\documentclass[aps,floatfix,pra,superscriptaddress,onecolumn,nofootinbib]{revtex4}
\usepackage{graphicx}
\usepackage{graphics}
\usepackage{amssymb}
\usepackage{amsmath}
\usepackage{latexsym}
\usepackage{hyperref}
\usepackage{ulem}
\usepackage{color}
\usepackage{subfig}

\newcommand{\ket}[1]{| #1\rangle}
\newcommand{\bra}[1]{\langle #1 |}

\def\ketc[#1]{\vert #1 \rangle}
\def\brac[#1]{\langle #1 \vert}

\newcommand{\expect}[1]{\langle{#1}\rangle}

\newcommand{\beq}{\begin{equation}}
\newcommand{\eeq}{\end{equation}}
\newcommand{\bqa}{\begin{eqnarray}}
\newcommand{\eqa}{\end{eqnarray}}
\newcommand{\nn}{\nonumber}

\newcommand{\erf}[1]{Eq.~(\ref{#1})}

\begin{document}

\title{Complete Characterization of Mixing Time for the Continuous Quantum
Walk on the Hypercube with Markovian Decoherence Model}
\author{Milosh Drezgich}
\affiliation{Department of Electrical Engeneering and Computer Science, University of
California, Berkeley, CA 94720, USA.}
\email{drezgic (at) eecs (dot) berkeley (dot) edu}
\author{Andrew P. Hines}
\affiliation{Pacific Institute of Theoretical Physics, Department of Physics and Astronomy, University of British Columbia,Vancouver BC, Canada V6T 1Z1.}
\author{Mohan Sarovar}
\affiliation{Department of Chemistry, University of California, Berkeley, CA 94720, USA.}
\author{Shankar Sastry}
\affiliation{Department of Electrical Engeneering and Computer Science, University of
California, Berkeley, CA 94720, USA.}

\begin{abstract}
The $n$-dimensional hypercube quantum random walk (QRW) is a particularily
appealing example of a quantum walk because it has a natural implementation
on a register on $n$ qubits. However, any real implementation will encounter
decoherence effects due to interactions with uncontrollable degrees of
freedom. We present a complete characterization of the mixing properties of
the hypercube QRW under a physically relevant Markovian decoherence model. In
the local decoherence model considered the non-unitary dynamics are modeled
as a sum of projections on individual qubits to an arbitrary direction on
the Bloch sphere. We prove that there is always classical (asymptotic)
mixing in this model and specify the conditions under which instantaneous
mixing \textit{always} exists. And we show that the latter mixing property,
as well as the classical mixing time, depend heavily on the exact
environmental interaction and its strength. Therefore, algorithmic
applications of the QRW on the hypercube, if they intend to employ mixing
properties, need to consider both the walk dynamics and the precise
decoherence model.
\end{abstract}

\maketitle

\section{Introduction, Previous Work, and Our Work}

Quantum walks \cite{Kem03} play a prominent role in the design of quantum
algorithms. Their distinction from classical random walks lies in their
potentially faster mixing and hitting times. 
The underlying dynamics of quantum walks can be either continuous or
discrete, and even though these two representations have some properties in
common, analysis usually demands different techniques and tools. Despite
their different origins, discrete and continuous time quantum walks can be
precisely related to each another \cite{Childs08}. 

The theoretical properties of quantum walks on general graphs have been
outlined in Ref. \cite{Aharonov}, and the remarkable result by Szegedy \cite%
{Szegedy} explains a general framework for the proper quantization of any
Markov chain algorithm. The dynamics of quantum walks have been analyzed,
for example, on the line \cite{Romanelli}, circle \cite{Solenov01},
hyperlattice \cite{PS00} and hypercube \cite{russell01}. Shenvi \textit{et.
al.} \cite{Shenvi} proved that the discrete quantum walk can be used in a
search algorithm and Ambainis used it for the best known algorithm for
element distinctness \cite{Ambainis}. The continuous-time quantum walk was
originally proposed by Childs, Farhi and Guttman \cite{FG98,Childs02} as an
algorithmic primitive. Childs \textit{et. al.} used a quantum walk to prove
the best known results for the separation between quantum and classical
query complexity \cite{Childs}, and a highly efficient algorithm for NAND
formula evaluation \cite{ChildsNAND}.

Central to the algorithmic application of both classical and quantum random
walks are their mixing characteristics. For a classical random walk that has
a unique steady state, the mixing time characterizes the convergence of the
walk to this steady state. In the quantum case, unitarity prevents the walk
from reaching a steady state. This has led to alternative notions of mixing
for quantum walks. One is to define an instantaneous mixing time, as the
first instant the probability distribution of the walker's location on the
graph is $\epsilon $-close to the uniform distribution. Another sensible
definition for the mixing time of a continuous quantum walk, although one
that is dependent on the initial state, is based on a limiting value of a
time-averaged probability distribution \cite{Aharonov}.

Quantum systems are very susceptible to imperfections and interactions with
their environment, both of which cause decoherence. Sufficient decoherence
can remove any potential benefits from the quantum dynamics. Investigations
to date have either used Markovian models for the environment, with the
environment monitoring the walker position or state of the `coin' driving
the walk \cite{KT03,BCA03a,BCA03b,BCA03c,Fed06,AR05,Fed06}; or imperfect
evolution \cite{MBSS02,RSA+04,ADSS07,ORD06,SBBH03,MPAD08}, such as broken
edges.

While decoherence is nominally the nemesis of quantum information
processing, it has been argued that decoherence can in fact be `useful' in
the context of quantum walks \cite{KT03}. Decoherence can be used to force a
quantum walk to mix to a uniform distribution, and in this paper we will
illustrate this for the quantum walk on the hypercube. Similar results have
been shown for quantum walks on the line \cite{KT03} and $N$-cycle \cite%
{MK07}\textbf{\ }using weak measurements of the walker's position. Additionally, Richter has provided elegant analyses of the general phenomena of quantum speedup of classical mixing processes under a restricted  decoherence model \cite{Ric06}. 

In ref. \cite{MPAD08}, Marquezino \textit{et al.} examined the discrete-time
quantum walk on the hypercube, and derived the limiting time-averaged
distribution in the coherent case (no decoherence). The mixing behavior,
both to this distribution and the uniform distribution, were considered for
a coherent walk as well as under the decohering effects of randomly breaking
links in the network. In the decoherent case, the walk was shown to approach
the uniform distribution. Interestingly there is an optimal decoherence rate
which provides the fastest convergence. A similar effect was found for the $%
N $-cycle in \cite{KT03}. Below we show that this can also be true for the
continuous-time quantum walk on the hypercube, but not always.

Hitting times\footnote{The hitting time is defined as the first time a given vertex, or set of
such, is reached.} and instantaneous mixing times for the continuous-time
version of the hypercube quantum walk with decoherence were calculated
recently by Alagic and Russell \cite{AR05}. Analytical results were derived
by exploiting the representation of the quantum walk on the hypercube as a
set of non-interacting qubits; a simple example of how spin networks may be
mapped to quantum walks \cite{HS07a}. We extend these results and provide a complete characterization of the
mixing time in terms of the decoherence intensity and projection direction.
In our case the projection direction of the decoherence operator can be in
any arbitrary direction in contrast to the previous analysis which was
restricted to decoherence in the computational basis $\{|0\rangle ,|1\rangle
\}.$ Moreover, we show that randomizing the direction of decoherence leads
to the depolarization channel that, regardless of the decoherence intensity,
shows universal mixing behavior.

The paper is organized as follows: we first describe the quantum walk on the
hypercube and its mapping to a register of qubits in section \ref{sec::HCQW}%
. Then in section \ref{sec::Dec_model} we introduce our decoherence model
and explicitly show that it is the continuous-time analogue of the standard
discrete-time projection model. The definitions of various mixing times are
introduced formally in section \ref{sec::Mix_times}. With our quantum walk
and decoherence model the hypercube quantum walk is separable as a product
over single qubits, and we need only consider single qubit dynamics, which
we discuss in section \ref{sec::single_qubit}. Analysis of simple channels,
randomized projections and numerical results are presented as subsections of
section \ref{sec::single_qubit}.

\section{Quantum Walk on the Hypercube}

\label{sec::HCQW}

Continuous-time quantum walks \cite{FG98} are defined over an undirected
graph with $N\equiv 2^{n}$ nodes, each labelled by an integer $i\in \lbrack
0,N-1]$. These walks can in general be described by the Hamiltonian 
\begin{eqnarray}
\hat{H}_{s} &=&\sum_{[ij]}\Delta _{ij}(t)\left( \hat{c}_{i}^{\dagger }\hat{c}%
_{j}+\hat{c}_{i}\hat{c}_{j}^{\dagger }\right) \;+\;\sum_{j}\epsilon _{j}(t)%
\hat{c}_{j}^{\dagger }\hat{c}_{j},  \notag  \label{Ham-qwa} \\
&\equiv &\sum_{[ij]}\Delta _{ij}(t)\left( |i\rangle \!\!\langle j|+|j\rangle
\!\!\langle i|\right) \;+\;\sum_{j}\epsilon _{j}(t)|j\rangle \!\!\langle j|,
\end{eqnarray}%
where each node $i$ corresponds to the quantum state, $|i\rangle =\hat{c}%
_{i}^{\dagger }|0\rangle $, and $[ij]$ denotes connected nodes $i$ and $j$.
The state $|i\rangle $ thus corresponds to a `particle' located at node $i$.
The first term in (\ref{Ham-qwa}) is a `hopping' term with amplitude $\Delta
_{ij}(t)$ between nodes $i$ and $j$; the second describes `on-site' node
energies $\epsilon _{j}(t)$. Both these terms can depend on time. One can
simplify (\ref{Ham-qwa}) by dropping all the on-site energies, and by making
all the internode hopping matrix elements the same, i.e., $\Delta
_{ij}\rightarrow \Delta ,\forall \{i,j\}$.

The structure of a hypercube is particularly appealing; it is an $n-$regular
graph that is the underlying model for many computational problems. The
nodes of this graph can be represented as basis vectors $|{v}\rangle \in \{|{%
1}\rangle ,|{0}\rangle \}^{\otimes n}$ in $\mathbb{C}^{2^{n}}$ dimensional
Hilbert space -- the same Hilbert space describing $n$ qubits. Each node is
labelled by a binary string representing a multi-qubit state, i.e. $|\vec{z}%
\rangle \equiv |z_{1}z_{2}\ldots z_{n}\rangle =|z_{1}\rangle \otimes
|z_{2}\rangle \otimes \ldots \otimes |z_{n}\rangle $, where the $%
z_{i}^{\prime }s$ are $0$ or $1$; pairs of nodes with a Hamming distance of $%
1$ (the number of bits that must be flipped to obtain one from the other)
are connected to give the hypercube. In this way $n$ qubits describe a
quantum walk over a $N$-dimensional hypercube, which takes place in
information space. This is described by the simple qubit Hamiltonian 
\begin{equation}
H=\Delta \sum_{i=1}^{n}\hat{\sigma}^i_{x},  \label{eq::Ham_qbit}
\end{equation}%
representing a set of non-interacting qubits, each evolving under the Pauli $\hat{%
\sigma}_{x} \equiv \left(\begin{array}{cc}0 & 1 \\1 & 0\end{array}\right)$ operator. Note that we do not scale the rate of Hamiltonian evolution with the number of qubits (e.g. $H \propto \Delta/n$ for some $\Delta$ independent of $n$) as was done in some prior work (e.g. Ref. \cite{russell01}). We believe this scaling is artificial and unnecessary for a quantum walk implemented using a qubit register.

In this case the unitary dynamics is trivially solvable; for the walker
initialized at the $\vec{z}=\vec{0}$ corner, the probability of being at
some site $\vec{z}$ is 
\begin{eqnarray}
P_{\vec{z}}(t)\equiv \langle \vec{z}|\varrho (t)|\vec{z}\rangle 
&=& \langle \vec{z}| ~ \bigotimes_{i=1}^n e^{-i\Delta \hat{\sigma}_x^i t} \ket{0}_i \bra{0} e^{i\Delta \hat{\sigma}_x^i t}~ |\vec{z}\rangle \nn \\
&=& \langle \vec{z}| ~ \bigotimes_{i=1}^n [ \cos(\Delta t)\ket{0} + \sin(\Delta t)\ket{1})(\cos(\Delta t)\bra{0} + \sin(\Delta t)\bra{1} ]  ~ |\vec{z}\rangle \nn \\
&=& \cos
^{2n_{0}}(\Delta t)\sin ^{2n_{1}}(\Delta t) \nn \\
 \label{Pz}
\end{eqnarray}%
where $n_{0}$ is the number of $0$'s, and $n_{1}$ the number of $1$'s
appearing in $\vec{z}$, and $\varrho (t)=e^{-iHt}|\vec{z}=0\rangle \langle 
\vec{z}=0|e^{iHt}$ is the density matrix of the qubit register at time $t$. The state of the qubit register is always pure, but we specify it with a density matrix to be consistent with what follows.

This mapping to a qubit model is not only useful conceptually, but it is
highly suggestive of a potential physical implementation of the hypercube
quantum walk. If the system is not closed and it is exposed to measurement
or an environment that it interacts with, the quantum dynamics are in general
far more complicated and we will examine this now.

\section{Decoherence Model}
\label{sec::Dec_model}
To study decoherence of quantum walks it is common to begin with a discrete-time quantum walk and model the decoherence as a sequence of weak measurements on the walker \cite{KT03,Ken06,BCA03a,BCA03b}.
When the result of the measurement is ignored (i.e. lost to the
environment), the nonunitary process is described by: 
\begin{equation}
\varrho _{n+1}=(1-p)\hat{U}\varrho _{n}\hat{U}^{\dagger
}+p\sum_{\alpha}M_{\alpha}(\hat{U}\varrho _{n}\hat{U}^{\dagger
})M_{\alpha}^{\dagger },  \label{eq::discrete-evo}
\end{equation}%
where the measurement is given by the POVM $\left\{ M_{\alpha}\right\} $,
such that $\sum_{\alpha}M_{\alpha}=1$ and $M_\alpha \geq 0$, occurring with
probability $p$ at each time step; $\hat{U}$ describes the unitary evolution
of the quantum walk. This model is equivalent to a memoryless environment,
unperturbed by the system. We will consider the continuous-time analogue of
this process, which we derive below.

\textbf{Claim:} The discrete-time weak measurement model of the system
dynamics, 
\begin{equation}
\varrho _{t+\tau }=(1-\gamma \tau )U_{\tau }\varrho _{t}U_{\tau }^{\dagger
}+\gamma \tau \sum_{\alpha}M_\alpha[U_{\tau }\varrho _{t}U_{\tau }^{\dagger }%
]M_\alpha^{\dagger }\,\,,  \label{eq::weak}
\end{equation}
is equivalent, in the limit $\tau \rightarrow 0$ to the master equation 
\begin{equation}
\dot{\varrho}(t)=-i[H,\varrho (t)]+\gamma \sum_{\alpha}\mathcal{D}%
[M_\alpha]\varrho (t)\,\,,  \label{eq::master}
\end{equation}%
if the measurement rate $\gamma $ is such that for a time-step of duration $%
\tau $, $p=\gamma \tau $, unitary evolution $\hat{U}_{\tau }=\exp \left( {%
-iH \tau }\right) $. Here the superoperator $\mathcal{D}[X]\varrho \equiv
X\varrho (t)X^{\dagger }-\frac{1}{2}\left( X^{\dagger }X\varrho (t)+\varrho
(t)X^{\dagger }X\right) $ for any operator $X$.

\textbf{Proof: \ } For small $\tau $ we expand the exponential to
first-order such that Eq. (\ref{eq::weak}) becomes 
\begin{eqnarray}
\varrho _{t+\tau } &\approx&(1-\gamma \tau )(\mathbf{1}-i\tau H)\varrho _{t}(%
\mathbf{1}+i\tau H)+\gamma \tau \sum_{\alpha}M_\alpha(\mathbf{1}-i\tau
H)\varrho _{t}(\mathbf{1}+i\tau H)M_\alpha^{\dagger }.
\end{eqnarray}
Now, dropping second order terms in the parameter $\tau$, 
\begin{eqnarray}
\varrho_{t+\tau } &\approx&\varrho _{t}+i\tau \varrho _{t}H-i\tau H\varrho _{t}-\gamma \tau \varrho
_{t}+\gamma \tau \sum_{\alpha}M_\alpha\varrho _{t}M_\alpha^{\dagger } \\
&=&\varrho _{t}+i\tau \varrho _{t}H-i\tau H\varrho _{t}\,+\gamma \tau
\sum_{\alpha}\left( M_\alpha\varrho _{t}M_\alpha^{\dagger }-\frac{1}{2}%
M_\alpha^{\dagger }M_\alpha\varrho _{t}-\frac{1}{2}\varrho
_{t}M_\alpha^{\dagger }M_\alpha\right) .
\end{eqnarray}%
Dividing by $\tau $ we obtain, 
\begin{equation}
\frac{\varrho _{t+\tau }-\varrho _{t}}{\tau }=-i[H,\varrho _{t}]\,+\gamma
\sum_{\alpha}\left( M_\alpha\varrho _{t}M_\alpha^{\dagger }-\frac{1}{2}%
[M_\alpha^{\dagger }M_\alpha\varrho _{t}-\varrho _{t}M_\alpha^{\dagger
}M_\alpha]\,\right) \,,
\end{equation}%
then taking the limit $\tau \rightarrow 0$, 
\begin{eqnarray}
\dot{\varrho}(t) &=&-i[H,\varrho (t)]\,+\gamma \sum_{\alpha}\left(
M_\alpha\varrho (t)M_\alpha^{\dagger }-\frac{1}{2}[M_\alpha^{\dagger
}M_\alpha\varrho (t)-\varrho (t)M_\alpha^{\dagger }M_\alpha]\,\right) \, \\
&=&-i[H,\varrho (t)]+\gamma \sum_{\alpha}\mathcal{D}[M_\alpha]\varrho
(t)\,\,\,,
\end{eqnarray}%
as claimed.$_{\blacksquare }$

The interaction between the system and the environment can be represented in
various ways, through the choice of the set of $\left\{ M_\alpha\right\} $.
In the quantum walk literature, the standard choice is the walker location,
i.e. $M_\alpha=|i\rangle \!\!\langle i|=\mathbb{P}_{i}$, the projectors onto
the graph node states. However, when the quantum walk on the hypercube is implemented via a set of qubits, as we describe here, position measurements corresponds to a computational basis
measurement of the state of every qubit simultaneously. This implies a
physically unrealistic, multi-qubit measurement/interaction with the
environment.

If the quantum walk were to be implemented using a qubit register, a more
physically realistic decoherence process is described by single-qubit
projective measurements. The obvious choice, as an analogue to the location
measurements, are measurements that are projections onto single qubit
computational basis states; this is the choice considered in \cite{AR05}. We
can generalize this to single-qubit projective measurements onto arbitrary
antipodal points on the Bloch sphere. We express this as: 
\begin{eqnarray}
\mathbb{P}_{0}(\overrightarrow{\mathbf{r}}) &=&\frac{\mathbb{I}+%
\overrightarrow{\mathbf{r}}\cdot \vec{\mathbf{\sigma}}}{2}\,\,\equiv \frac{%
\mathbb{I}+(\sin \theta \cos \varphi) \sigma _{x}+(\sin \theta \sin \varphi)
\sigma _{y}+(\cos \theta) \sigma_{z}}{2}\in \mathbb{C}^{2\times 2}\,\,, 
\notag \\
\mathbb{P}_{1}(\overrightarrow{\mathbf{r}}) &=&\frac{\mathbb{I}-%
\overrightarrow{\mathbf{r}}\cdot \vec{\mathbf{\sigma}}}{2}\,\,\equiv \frac{%
\mathbb{I}-(\sin \theta \cos \varphi) \sigma _{x}-(\sin \theta \sin \varphi)
\sigma _{y}-(\cos \theta) \sigma_{z}}{2}\in \mathbb{C}^{2\times 2}\,\,,
\end{eqnarray}%
where $\mathbb{I}$ is the two-dimensional identity matrix, $\overrightarrow{%
\mathbf{r}}=(r_{x},r_{y},r_{z})=(\sin \theta \cos \varphi ,\sin \theta \sin
\varphi ,\cos \theta )$, for $\theta \in \lbrack 0,\pi ]$, $\varphi \in
\lbrack 0,2\pi ]$, defines what we refer to as the \textquotedblleft
decoherence axis\textquotedblright\ or measurement projection direction. By $%
\vec{\mathbf{\sigma}}=(\sigma _{x},\sigma _{y},\sigma _{z})$ we denote the
three dimensional matrix vector composed of the three nontrivial Pauli
matrices. This decoherence model is a generalization of what has been
referred to as the \textit{subspace projection} decoherence model \cite%
{Strauch08}.

We consider the continuous quantum walk described by the master equation Eq. (\ref%
{eq::master}), with a Hamiltonian given by Eq. (\ref{eq::Ham_qbit}) and measurement operators given by 
\begin{equation}
M^{k}_{\alpha }=\mathbb{I}\otimes \ldots \otimes \mathbb{P}_{\alpha }(%
\overrightarrow{\mathbf{r}})\otimes \ldots \otimes \mathbb{I},
\label{eq:povm_elems}
\end{equation}%
where $\alpha =\left\{ 0,1\right\} $ and the projector is on the $k^{th}$
qubit. Note that $\sum_\alpha M^k_\alpha = \mathbb{I}^{\otimes n}$. The
qubit register evolution equation can then be written as: 
\begin{equation}
\dot{\varrho}(t)=\sum_{k=1}^{n}-i\Delta \left[ \sigma ^{k}_{x},\varrho (t)%
\right] + \gamma\sum_{k=1}^{n}\sum_{\alpha=0}^1 \mathcal{D}\left[ M_\alpha^k %
\right] \varrho (t).  \label{eq:tovec}
\end{equation}
This can alternatively be written as: 
\begin{equation}
\dot{\varrho}(t)=\sum_{k=1}^{n}-i\Delta \left[ \sigma ^{k}_{x},\varrho (t)%
\right] +\frac{\gamma }{2}\sum_{k=1}^{n}\mathcal{D}\left[ \overrightarrow{%
\mathbf{r}}\cdot \vec{\mathbf{\sigma}}_{k}\right] \varrho (t),
\label{eq:master_multiqubit}
\end{equation}%
where the sum is over qubits. None of the qubits are interacting and
therefore this master equation has a separability property that allows us to
treat the dynamics as $n$ single qubit density matrices undergoing the
evolution: 
\begin{equation}
\dot{\rho_k}(t)=-i\Delta \left[ \sigma_{x},\rho_k (t)\right] +\frac{\gamma}{2%
} \mathcal{D}\left[ \overrightarrow{\mathbf{r}}\cdot \vec{\mathbf{\sigma}}%
\right] \rho_k (t),  \label{eq::master2}
\end{equation}
and $\varrho = \bigotimes_{k=1}^n \rho_k$. See the appendix for an explicit
derivation of this. This property allows one to analyze the full system
dynamics by looking at single qubit dynamics, since the dynamics of the
system is just the tensor product of the dynamics of individual, non-interacting
qubits. Furthermore, the evolution equation for $\rho_k$ is the same for all 
$k$, and so we will drop the subscript when referring to single qubit
dynamics.

We will investigate how changes in the single qubit dynamics affect
properties of the quantum walk on the hypercube. 
After formally defining the mixing time of a quantum walk in the next
section, we provide a complete characterization of the mixing time in terms
of the quantum channel in Eq. (\ref{eq::master2}), i.e. dependence on the
physical rates and the direction of the decoherence axis, $\overrightarrow{%
\mathbf{r}}$ \footnote{%
Note that varying $\overrightarrow{\mathbf{r}}$ is equivalent to changing
the basis in which node states are encoded while the decoherence axis is
kept fixed.}.

\section{Mixing Time}

\label{sec::Mix_times}

To identify the physical quantities of interest, we return to a principal
motivation for considering random walks (quantum or classical). In computer
science, the most efficient solution to many problems is given by a
probabilistic algorithm, where the correct answer is attained with
high-probability if the space of solutions is sampled with a well-chosen
sampling distribution. Generating the correct sampling distribution is often
a matter of mapping the uniform distribution into the desired one, and
therefore generating a truly uniform distribution is an important problem.

There are several different definitions in the literature that have been
used as a measure of mixing time. For completeness will briefly list them
all here. For our purposes the notions of instantaneous mixing and classical
mixing will be the most important ones.

\textit{Instantaneous mixing} is defined as the first time instant at which
the probability distribution of the walker's position is sufficiently close
to uniform distribution:\ 
\begin{equation}
M_{inst,\varepsilon }=\min \{t~|~||P(x,t)-P_{u}||_{tv}<\varepsilon \}
\end{equation}%
where $P(x,t)$ is the probability of obtaining element $x\in \mathcal{X}$ ($%
\mathcal{X}$ is the space of events we are sampling from, which in the case
of random walks is the space of the walker location parameter) at time $t$,
and $P_{u}$ is the uniform distribution over $\mathcal{X}$. $||\cdot ||_{tv}$
is the \textit{total variation} distance over probability distributions (we
will restrict our attention to finite sample spaces). This definition is
mostly used in idealized continuous quantum random walks where no
decoherence effects are present. Although formally present by this
definition, mixing in continuous quantum random walks without decoherence is
only an instantaneous phenomenon. The ability to harness this instantaneous
mixing is still questionable.

\textit{Average mixing} is based on the time-averaged probability
distribution, that even for unitary quantum walks is shown to converge \cite%
{Aharonov}. In the continuous-time case the time-averaged probability
distribution for the state $x$ is defined as: 
\begin{equation}
\bar{P}(x,\tau )=\frac{1}{\tau }\int_{0}^{\tau }P(x,t)dt\,,\,\,
\end{equation}%
We can define the corresponding time-averaged mixing as: 
\begin{equation}
M_{avg,\varepsilon }=\min \{T~|~\forall \tau >T:||\bar{P}(x,\tau
)-P_{u}||_{tv}<\varepsilon \}.  \label{eq::AvgMix}
\end{equation}%
This time-averaged distribution can be sampled from by selecting $t$
uniformly in $[0,T]$, running the quantum walk for time $t$, and measuring
the walker position.

\textit{Classical (asymptotic) mixing} is the quantity originally used in
classical random walks and it defines mixing as the time after which the
register's distribution is desirably close to uniform: 
\begin{equation}
M_{class,\varepsilon }=\min \{T~|~\forall
t>T:||P(x,t)-P_{u}||_{tv}<\varepsilon \}
\end{equation}%
This definition characterizes the time it takes for the probability of
finding the walker at a particular location to be distributed uniformly
across the entire sample space, $\mathcal{X}$.

The mixing time is a well defined quantity for classical random walks
because there exists a stationary distribution for classical random walks
over any connected, non-bipartite graph \cite{Lov1996}, however for
continuous quantum walks this is not necessarily the case; unitary dynamics
means the probability distribution over the graph nodes oscillates, and
therefore never converges to the uniform distribution. We shall return to
this issue below. But first, let us examine the total variational distance
in the context of a quantum walk on a hypercube.

For a hypercube quantum walk implemented using qubits, the walker location
is encoded into the value of the qubit register in the computational basis.
Therefore, the sample space in this case is the space of binary strings of
length $n$, and the probability of measuring any register value (walker
\textquotedblleft location"), $\vec{z}$ is: 
\begin{equation}
P(\underline{x},t)=(1-p_{0}(t))^{k}p_{0}(t)^{n-k}
\end{equation}%
where $k$ is the Hamming weight of the binary string $\vec{z}$ (i.e. number
of ones in $\vec{z}$), and $p_{0}(t)$ is the probability of a qubit value
being $0$ at time $t$. Note that: 
\begin{equation*}
p_{0}(t)=\frac{1+\langle {\sigma _{z}}(t)\rangle }{2}
\end{equation*}%
where $\langle A\left( t\right) \rangle \equiv \text{tr}(A\rho (t))$ for any
operator $A$. The total variational distance in this case is: 
\begin{eqnarray}
||P(\vec{z},t)-P_{u}||_{tv} &=&\sum_{\vec{z}}|P(\vec{z},t)-\frac{1}{2^{n}}| 
\notag \\
&=&\sum_{k}{\binom{n}{k}}|(1-p_{0}(t))^{k}p_{0}(t)^{n-k}-\frac{1}{2^{n}}|
\end{eqnarray}%
We can bound this variational distance using the \textit{Hellinger distance}
on distributions \cite{AR05}. The Hellinger distance, for two distributions $%
\mathcal{P}(x)$ and $\mathcal{Q}(x)$ (both defined over the same sample
space $\mathcal{X}$), is defined as: 
\begin{equation}
H(\mathcal{P},\mathcal{Q})^{2}\equiv\frac{1}{2}\sum_{x}[\sqrt{\mathcal{P}(x)}-\sqrt{\mathcal{%
Q}(x)}]^{2}=1-\sum_{x}\sqrt{\mathcal{P}(x)\mathcal{Q}(x)}
\end{equation}%
Its usefulness for us comes from its relation to the total variational
distance: 
\begin{equation}
||\mathcal{P}-\mathcal{Q}||_{tv}\leq 2H(\mathcal{P},\mathcal{Q})\leq 2||%
\mathcal{P}-\mathcal{Q}||_{tv}^{1/2}
\end{equation}%
Therefore in our case, 
\begin{eqnarray}
||P(\vec{z},t)-P_{u}||_{tv}^{2} &\leq &4H(P(\vec{z},t),P_{u})^{2}  \notag \\
&=&4-4\sum_{\vec{z}}\sqrt{P(\vec{z},t)\frac{1}{2^{n}}}=4-4\sum_{k}{\binom{n}{%
k}}\sqrt{\frac{(1-p_{0}(t))^{k}p_{0}(t)^{n-k}}{2^{n}}}  \notag \\
&=&4-4\left( \sqrt{\frac{1-p_{0}(t)}{2}}+\sqrt{\frac{p_{0}(t)}{2}}\right)
^{n}  \notag \\
&=&4\left[ 1-\frac{1}{2^{n}}\left( \sqrt{1-\langle {\sigma _{z}}(t)\rangle }+%
\sqrt{1+\langle {\sigma _{z}}(t)\rangle }\right) ^{n}\right] 
\label{eq::bound}
\end{eqnarray}%
Note that this is a positive quantity because we only take the positive
branch of the square roots, and it attains its minimum value of zero when $%
\langle {\sigma _{z}}(t)\rangle =0$. Furthermore, from this bound we see
that the variational distance for the distribution of the register of $n$
qubits (from the uniform distribution) is small exactly when the variational
distance for the distribution of a single qubit is small. In fact, when $%
\langle {\sigma _{z}}\rangle $ is small, we can expand the square roots to
second order in this quantity to get: 
\begin{equation}
||P(\vec{z},t)-P_{u}||_{tv}^{2}\leq 4-4\left( 1-\frac{\langle {\sigma _{z}}%
(t)\rangle ^{2}}{8}\right) ^{n}  \label{eq::tv_distance}
\end{equation}%
Given this fact, we will concentrate on the distribution for a single qubit
in the following, and appeal to the fact that the variational distance for
the entire register from the uniform distribution is small precisely when $%
\langle {\sigma _{z}}(t)\rangle ^{2}$ is small for a single qubit.

Finally, although our intention in this work is not to characterize the mixing behavior of the random walk as a function of $n$, the number of qubits, we note that the above inequality allows us to easily place bounds on this behavior. Consider the time at which the total variational distance of the register of qubits is at most $\varepsilon$; the above bound tells us that this is when:
\bqa
n\ln \left( 1-\frac{\langle {\sigma _{z}}(t)\rangle ^{2}}{8}\right) = \ln\left( 1-\frac{\varepsilon}{4}\right)
\label{eq:mix_cond}
\eqa
Expanding both logs in the small parameters $\varepsilon$ and $\sigma _{z}(t)\rangle ^{2}/{8}$ (because $\expect{\sigma _{z}(t)}$ will be small whenever this equality is fulfilled) we get $\expect{\sigma _{z}(t)}  = \sqrt{\frac{2\varepsilon}{n}}$. Now, we shall show in the following sections that the behavior of the single qubit $\expect{\sigma_{z}(t)}$ as a function of time always has an exponentially decaying envelope when there is non-zero decoherence ($\gamma \neq 0$). Therefore, to bound the time for asymptotic mixing we can substitute $\expect{\sigma_{z}(t)} = Ce^{-r t}$ into \erf{eq:mix_cond}, and get the following bound on the behavior of register: $t_{\textrm{mix}} = \mathcal{O}(\ln n)$. This scaling behavior has been noted elsewhere and will not be the subject of study in this work. Instead, we will investigate how the mixing of a single qubit is affected by the physical parameters in the problem, particularly the decoherence parameters. In effect, we will study how the values of constants $C$ and $r$ in the exponential envelope above are determined by the physical parameters.

\section{Single-qubit dynamics}

\label{sec::single_qubit}

In order to solve for the single qubit dynamics we use the following
parametrization of the single qubit density operator, 
\begin{equation}
\rho (t)\,=\frac{1}{2}\left( 
\begin{array}{cc}
1+\langle \sigma _{z}(t)\rangle & \langle \sigma _{x}(t)\rangle -i\langle
\sigma _{y}(t)\rangle \\ 
\langle \sigma _{x}(t)\rangle +i\langle \sigma _{y}(t)\rangle & 1-\langle
\sigma _{z}(t)\rangle%
\end{array}%
\right) \,\,,  \label{eq::param}
\end{equation}%
This operator is completely described by the Bloch vector $\langle \vec{%
\sigma}\rangle =(\langle \sigma _{x}\rangle ,\langle \sigma _{y}\rangle
,\langle \sigma _{z}\rangle )\equiv (x,y,z)$. We can then derive the matrix
equation for the Bloch vector of a single qubit when the evolution is given
by Eq.~(\ref{eq::master2}), repeated here for clarity:

\begin{equation}
\dot{\rho}(t)=-i\Delta \left[ \sigma _{x},\rho (t)\right] -\frac{\gamma }{2}%
\rho(t)+\frac{\gamma }{2}\left( \mathbf{r}\cdot \mathbf{\sigma }\right)
\rho(t)\left( \mathbf{r}\cdot \mathbf{\sigma }\right) ^{\dagger }.
\end{equation}%
Multiplying this equation by each of the Pauli matrices and taking the trace
we obtain the $3\times 3$ parametrized linear system equation:

\begin{equation}
\langle {\overrightarrow{\dot{\mathbf{\sigma}}}}\left( {t}\right) \rangle
\equiv \frac{d}{dt} \text{tr}(\overrightarrow{\mathbf{\sigma}} \rho(t) ) = 
\text{tr}( \overrightarrow{\mathbf{\sigma}} \dot{\rho}(t) ) = \mathbf{A}%
\langle {\overrightarrow{\mathbf{\sigma} }}\left( {t}\right) \rangle \,,
\label{eq::lin_eqn}
\end{equation}%
where, 
\begin{equation}
\mathbf{A}=\left( 
\begin{array}{ccc}
\gamma (r_{x}^{2}-1) & \gamma r_{x}r_{y} & \gamma r_{x}r_{z} \\ 
\gamma r_{x}r_{y} & \gamma (r_{y}^{2}-1) & \gamma r_{y}r_{z}-2\Delta \\ 
\gamma r_{x}r_{z} & \gamma r_{y}r_{z}+2\Delta & \gamma (r_{z}^{2}-1)%
\end{array}%
\right) .
\end{equation}%
The solution of this system\ is $\langle {\overrightarrow{\mathbf{\sigma} }}%
\left( t\right) \rangle =\exp (\mathbf{A}t)\langle {\overrightarrow{\mathbf{%
\sigma} }}\left( 0\right) \rangle ,$ where $\langle {\overrightarrow{\mathbf{%
\sigma} }}\left( 0\right) \rangle $ is the Bloch vector of the initial state 
$|0\rangle $. The dynamics of the qubit, and in turn the entire register
implementing the quantum walk, is completely determined by the properties
the matrix $\mathbf{A}$. We now examine the key properties of this matrix.

Firstly, by the Routh-Hurwitz criterion \cite{Stengel}, the matrix $\mathbf{A%
}$ has eigenvalues that lie in the left half of the complex plane for
positive $\gamma, \Delta$ and all decoherence axes \textit{except} for when $%
r_x=1$. This singular case represents a channel where the Hamiltonian
dynamics and the decoherence dynamics commute. This case is easy to solve
for explicitly (we present the solution in \ref{sec:examples}) and the
dynamics for it are fairly uninteresting. For all other parameter regimes,
the Routh-Hurwitz criterion tells us that Eq. (\ref{eq::lin_eqn}) is a
strictly stable system.

The eigenvalues of A can be determined from its characteristic equation: 
\begin{equation}
\lambda ^{3}+2\gamma \lambda ^{2}+(\gamma ^{2}+4\Delta ^{2})\lambda +4\gamma
\Delta ^{2}(1-r_{x}^{2})=0
\end{equation}%
Interestingly the eigenvalues depend only on $r_{x},\gamma ,\Delta $. The
most convenient form for the solutions to this cubic equation can be found
by mapping the equation into a third order Chebyshev polynomial and using
the Chebyshev cube root. Then the solution can be written in closed form as: 
\begin{eqnarray}
\lambda _{1} &=&2\sqrt{\frac{\gamma ^{2}-12\Delta ^{2}}{9}}\cos (\frac{1}{3}%
\arccos (m))-\frac{2\gamma }{3}  \nn \\
\lambda _{2} &=&-2\sqrt{\frac{\gamma ^{2}-12\Delta ^{2}}{9}}\cos (\frac{1}{3}%
\arccos (-m))-\frac{2\gamma }{3}  \notag \\
\lambda _{3} &=&-\lambda _{1}-\lambda _{2}-2\gamma ,
\label{eq:roots}
\end{eqnarray}%
where 
\begin{equation}
m=\frac{\gamma }{(\gamma ^{2}-12\Delta ^{2})^{3/2}}(\gamma ^{2}+18\Delta
^{2}(3r_{x}^{2}-1))).
\end{equation}%
These eigenvalues fall into one of two classes, depending on the values of $%
\gamma ,\Delta ,$and $\overrightarrow{\mathbf{r}}$: one real and two
imaginary eigenvalues, or three (with possible repetitions) real eigenvalues.

$\mathbf{A}$ is not a symmetric matrix and is therefore not generally
diagonalizable. $\mathbf{A}$ could be not diagonalizable if it has repeated
eigenvalues (i.e. a degenerate eigenspace). We will see below when we
perform a more detailed analysis of the eigenvalues of $\mathbf{A}$ that
this only occurs in a vanishingly small parameter range which will not be of
interest to us. Therefore we will effectively treat $\mathbf{A}$ as
diagonalizable.

Given these properties of the matrix $\mathbf{A}$, let us return to the
mixing behavior of a single qubit under the dynamics given by Eq. (\ref%
{eq::lin_eqn}). 
Assuming that the initial state of the whole register of the system is in
the state $|0\rangle ^{\otimes n}$ at $t=0$ the $\langle {\sigma _{z}}%
(t)\rangle $ component of each individual qubit is:%
\begin{equation}
\langle {\sigma _{z}}(t)\rangle =(0~0~1)e^{\mathbf{A}t}\langle {%
\overrightarrow{\mathbf{\sigma} }}(0)\rangle =(0~0~1)e^{\mathbf{A}%
t}(0~0~1)^{T}=\left( e^{\mathbf{A}t}\right) _{33}
\end{equation}%
For diagonalizable $\mathbf{A}$, with eigenvalues $\lambda _{j}$, $e^{%
\mathbf{A}t}$ can be written as: 
\begin{equation}
e^{\mathbf{A}t}=\sum_{j=1}^{3}e^{\lambda _{j}t}\prod_{k\neq j}\frac{\mathbf{A%
}-\lambda _{k}\mathbb{I}}{\lambda _{j}-\lambda _{k}},
\end{equation}%
Using this expansion, 
\begin{equation}
\langle {\sigma _{z}}(t)\rangle =\sum_{\pi (\lambda _{1},\lambda
_{2},\lambda _{3})}\frac{e^{\lambda_{\pi _{1}}t}}{(\lambda _{\pi
_{_{1}}}-\lambda _{\pi _{_{2}}})(\lambda _{\pi _{_{1}}}-\lambda _{\pi
_{_{3}}})}\left[ -4\Delta ^{2}+\gamma ^{2}(1-r_{z}^{2})+(\lambda _{\pi
_{_{2}}}+\lambda _{\pi _{_{3}}})\gamma (1-r_{z}^{2})+\lambda _{\pi
_{_{2}}}\lambda _{\pi _{_{3}}}\right]  \label{eq::z_t}
\end{equation}%
where $\pi (\lambda _{1,}\lambda _{2},\lambda _{3})$ denotes the three
cyclic permutations of $\left( \lambda _{1},\lambda _{2},\lambda _{3}\right) 
$. From this expression, we can see that $\langle {\sigma _{z}}(t)\rangle $
depends on all the \textit{free} parameters in the system: $%
r_{x},r_{z},\gamma ,\Delta $ (even though the eigenvalues only depend on $%
r_{x},\gamma $ and $\Delta $).

This expression for $\langle {\sigma _{z}}\rangle (t)$ tells us something
crucial about the QRW. The exponential envelopes $e^{\lambda_{\pi _{1}}t}$,
and the fact that $\mathbf{A}$ is a strictly stable matrix (has eigenvalues
in the left half of the complex plane), imply that $\langle {\sigma _{z}}%
\rangle \xrightarrow{t\rightarrow
\infty}0$. Hence, by Eq. (\ref{eq::tv_distance}), the limiting distribution
for the quantum walk for all $\overrightarrow{\mathbf{r}}$ and $\gamma >0$
is the uniform distribution. Thus, the decoherence ensures that the random
walk mixes to uniform given sufficient time. 

Finally, we note that although the above convergence argument was for the
particular initial state, $|0\rangle ^{\otimes n}$, the strict stability of
the dynamical matrix $\mathbf{A}$ (for all parameters except $r_x=1$) allows
us to state more generally that there is an initial state independent steady
state $\langle \overrightarrow{\mathbf{\sigma}} \rangle \xrightarrow{t%
\rightarrow\infty}\mathbf{0}$, which of course implies $\langle {\sigma _{z}}%
\rangle \xrightarrow{t\rightarrow \infty}0$.

\subsection{Special Cases: Simple Channels}

\label{sec:examples}

In this section, to illustrate the utility of our approach we exactly solve
the dynamics for several simple single-qubit channels.

For $r_{x}=0$, the dynamics is described by the matrix 
\begin{equation}
\mathbf{A}=\left( 
\begin{array}{ccc}
-\gamma & 0 & 0 \\ 
0 & \gamma (r_{y}^{2}-1) & \gamma r_{y}r_{z}-2\Delta \\ 
0 & \gamma r_{y}r_{z}+2\Delta & \gamma (r_{z}^{2}-1)%
\end{array}%
\right) .
\end{equation}%
So $\langle \sigma _{x}(t)\rangle =e^{-\gamma t}\langle \sigma
_{x}(0)\rangle $, and we have a pair of coupled linear differential
equations describing the motion in the $y$-$z$ plane. The motion in this
plane is analogous to a damped simple harmonic oscillator with natural
frequency $2\Delta $ and damping rate $\gamma $; for $\gamma <4\Delta $ the
system is underdamped, resulting in decaying oscillations around the origin,
while in the overdamped regime, $\gamma >4\Delta $, we see exponential
decay. In analyzing the mixing properties of the hypercube quantum walk, we
need only consider the behavior of the $z$-component of the Bloch vector.
When the initial state is $\langle \overrightarrow{\mathbf{\sigma} }%
(0)\rangle = (0,~0,~1)^T$, in the underdamped case, $\gamma <4\Delta $, we
have, 
\begin{equation}
\langle \sigma _{z}(t) \rangle =e^{-\gamma t/2}\left( \cos \omega t+\frac{%
\gamma (2r_{z}^{2}-1)}{2\omega }\sin \omega t\right) .
\label{eq::Zvar_nz_under}
\end{equation}%
where $\omega \equiv \sqrt{|\gamma^2 - 16\Delta^2|}/2$. In the overdamped
regime, $\gamma >4\Delta $, 
\begin{equation}
\langle \sigma _{z}(t) \rangle=\frac{1}{\sqrt{\gamma ^{2}-16\Delta ^{2}}}%
\left( (\lambda _{+}+\gamma r_{z}^{2})e^{\lambda _{+}t}-(\lambda _{-}+\gamma
r_{z}^{2})e^{\lambda _{-}t}\right) ,  \label{eq::Zvar_nz_over}
\end{equation}%
where $\lambda _{\pm }=(-\gamma \pm \sqrt{\gamma ^{2}-16\Delta ^{2}})/2$.
Finally, in the critically damped case $\gamma =4\Delta $, $\langle \sigma
_{z}(t) \rangle=\left( 2\Delta (2r_{z}^{2}-1)t+1\right) e^{-2\Delta t}$.

This result applies to two standard, single-qubit decoherence channels \cite%
{Chuang00}; the phase-flip, or dephasing, channel where $\overrightarrow{%
\mathbf{r}}=(0,0,1)$ and the bit-phase-flip channel $\overrightarrow{\mathbf{%
r}}=(0,1,0)$.

In the opposite case, where $\overrightarrow{\mathbf{r}}=(1,0,0)$ (the
bit-flip channel), 
\begin{equation}
\mathbf{A}=\left( 
\begin{array}{ccc}
0 & 0 & 0 \\ 
0 & -\gamma & -2\Delta \\ 
0 & 2\Delta & -\gamma%
\end{array}%
\right) ,
\end{equation}%
and the solution for $z(t)$ ,with initial condition $\langle \overrightarrow{%
\mathbf{\sigma} }(0)\rangle = (0,~0,~1)^T,$ is simply 
\begin{equation}
\langle \sigma _{z}(t) \rangle=e^{-\gamma t}\cos 2\Delta t.
\label{eq::Xbit-dep}
\end{equation}

Finally we examine the depolarizing channel \cite{Chuang00}, which although
is not an instance of the class of channels described by (\ref{eq::master2}%
), can be obtained by a randomization of such channels (see the Appendix for
the detailed derivation of this). The depolarization channel corresponds to
a randomization of the decoherence axis, $\overrightarrow{\mathbf{r}}$, and
is described by a dynamical matrix: 
\begin{equation}
\mathbf{A}=\left( 
\begin{array}{ccc}
-\frac{2\gamma }{3} & 0 & 0 \\ 
0 & -\frac{2\gamma }{3} & -2\Delta  \\ 
0 & 2\Delta  & -\frac{2\gamma }{3}%
\end{array}%
\right) .
\end{equation}%
The solution for the z-component is: 
\begin{equation}
\langle \sigma _{z}(t)\rangle =e^{-\frac{2\gamma }{3}t}\cos 2\Delta t.
\label{eq::Zbit-dep}
\end{equation}

\subsection{The Complete Classification of Mixing Behavior}

\label{subsec:eig} For dynamics under a general single qubit channel, the
convergence of the $\langle {\sigma _{z}}\left( t\right) \rangle $ to the
limiting value $0$ can be essentially of two different types, depending on
the eigenvalues of $\mathbf{A}$: exponential decay or dampened oscillator
decay. We now characterize the type of decay in terms of the physical
parameters of the qubit model, and then use the decay types to make
conclusions about the mixing properties of the quantum walk.

As we have already seen the characteristic equation for the matrix $\mathbf{A%
}$ is: 
\begin{equation}
\lambda ^{3}+2\gamma \lambda ^{2}+(\gamma ^{2}+4\Delta ^{2})\lambda +4\gamma
\Delta ^{2}\eta =0,  \label{eq:char_eqn}
\end{equation}%
denoting {$\eta \equiv 1-r_{x}^{2}.$ } For a cubic equation with real
coefficients the solutions are either: (1) three real, possibly repeated,
roots, or (2) one real root and two complex conjugate roots. According to
Eq. (\ref{eq::z_t}) these two classes of roots give rise to two
different types of convergence of $\langle \sigma_z \rangle
(t) $ to zero: exponential decay or dampened oscillator decay. To determine
the parameter dependence of these two types of convergence behavior, we can
use the discriminant of (\ref{eq:char_eqn}): 
\begin{equation}
\Lambda (\eta )=432\gamma ^{2}\Delta ^{4}\eta ^{2}-(16\gamma ^{4}\Delta
^{2}+576\gamma ^{2}\Delta ^{4})\eta +(4(\gamma ^{2}+4\Delta
^{2})^{3}-4\gamma ^{2}(\gamma ^{2}+4\Delta ^{2})^{2}).
\end{equation}%
There are two fundamentally distinct parameter regions:

\begin{itemize}
\item[\textit{(i)}] \textit{(Zeno region)} $\Lambda (\eta )\leq 0\Rightarrow
3\,$ real roots, that are distinct unless,

$\Lambda (\eta )=0,$ in which case either two or all three are repeated,

\item[\textit{(ii)}] \textit{(no-Zeno region)} $\Lambda (\eta )>0\Rightarrow
1~$real root and$~2~$complex conjugate roots .{\ }
\end{itemize}

The reason for the names for the two regions, \textit{Zeno} and \textit{%
no-Zeno}, will become clear when we examine the mixing behavior of random
walks with dynamics prescribed by a dynamical matrix $\mathbf{A}$ that lies
in one of the above regions. First, let us define the border between the
Zeno and no-Zeno regions in terms of the parameters $r_{x}, \gamma$ and $%
\Delta$. This border is defined by the values where the discriminant equals
zero:\ {\ 
\begin{equation}
\Lambda (\eta )=0\Leftrightarrow \eta =\frac{2}{3}+\frac{\gamma ^{2}}{%
54\Delta ^{2}}\pm \frac{\sqrt{\left( \gamma ^{2}-12\Delta ^{2}\right) ^{3}}}{%
54\gamma \Delta ^{2}};  \label{border}
\end{equation}%
}

When $\gamma <\sqrt{12}\Delta $ the equation $\Lambda (\eta )=0$ has no real
solutions, and therefore we cannot have repeated roots in this parameter
range. On the other hand, when $\gamma \geq \sqrt{12}\Delta $ we have two
real values for $\eta $ that define the upper and lower boundary of the
no-Zeno parameter region. Since {$\eta \equiv 1-r_{x}^{2}$ we get the
following expressions: 
\begin{equation}
-\text{ for}~~\sqrt{12}\Delta \leq \gamma \leq 4\Delta, ~~\text{repeated
eigenvalues when}~~r_{x}=\pm \sqrt{\frac{1}{3}-\frac{\gamma ^{2}}{54\Delta
^{2}}\pm \frac{\sqrt{\left( \gamma ^{2}-12\Delta ^{2}\right) ^{3}}}{54\gamma
\Delta ^{2}};}  \label{aregion}
\end{equation}%
}

{%
\begin{equation}
-\text{ for\ }\gamma >4\Delta, ~\text{repeated eigenvalues when}~~r_{x}=\pm 
\sqrt{\frac{1}{3}-\frac{\gamma ^{2}}{54\Delta ^{2}}+\frac{\sqrt{\left(
\gamma ^{2}-12\Delta ^{2}\right) ^{3}}}{54\gamma \Delta ^{2}}}.
\end{equation}%
}

\begin{figure}[t!]
\centering 
\subfloat[~Phase diagram for eigenvalues of the dynamical matrix
$\mathbf{A}$. The red (shaded) region indicates where eigenvalues are purely
real, or what is referred to as the Zeno-region in the main
text.]
{ 
\label{fig:eig_phasediag:a} 
\includegraphics[width=0.45\linewidth]{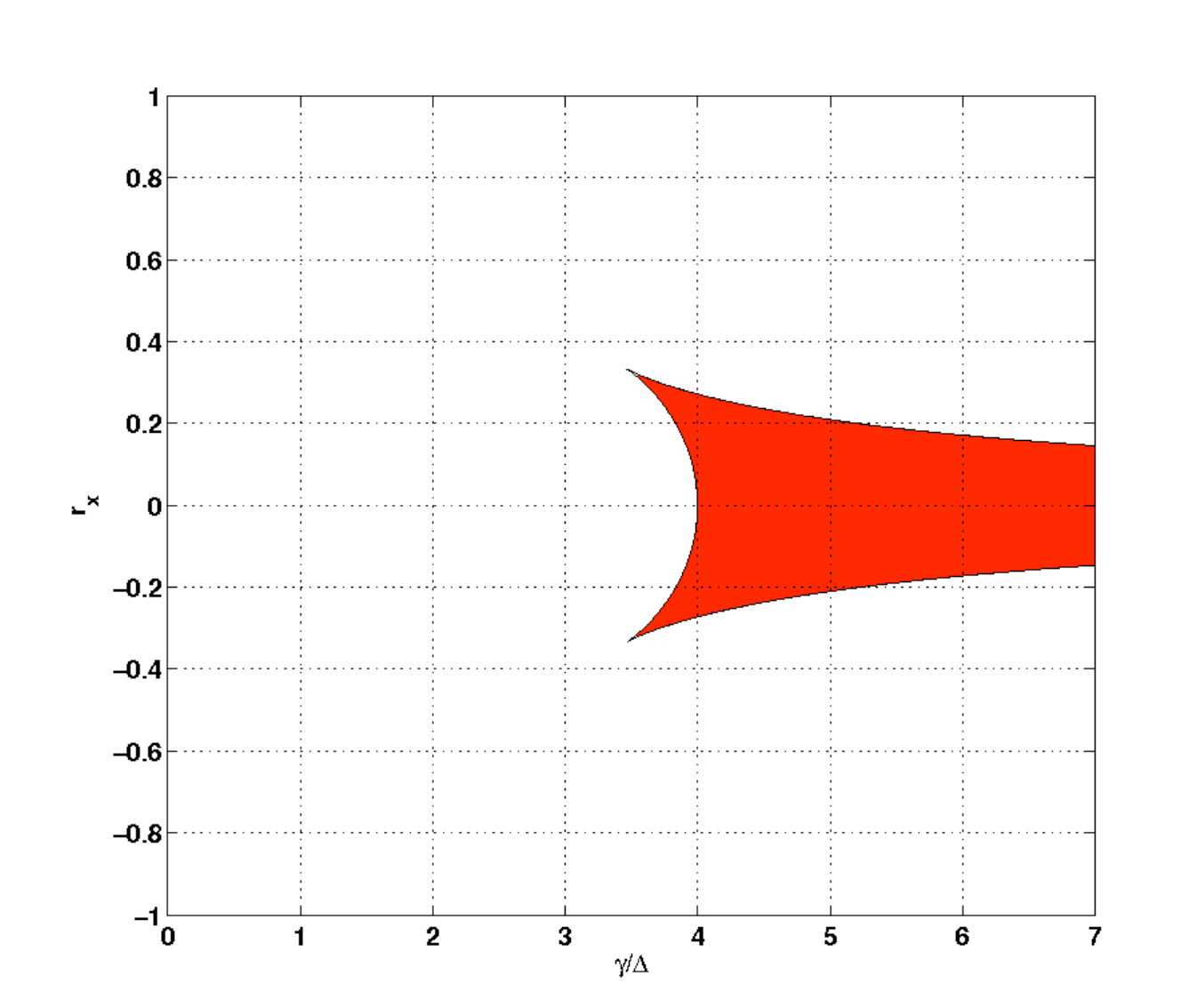}} 
\hspace{0.01\linewidth} 
\subfloat[~$\mathbf{r}$-space diagram of eigenvalue types for matrix $\mathbf{A}$ when $\gamma/\Delta = \sqrt{15}$.]{ 
\label{fig:eig_phasediag:b}
\includegraphics[width=0.25\linewidth]{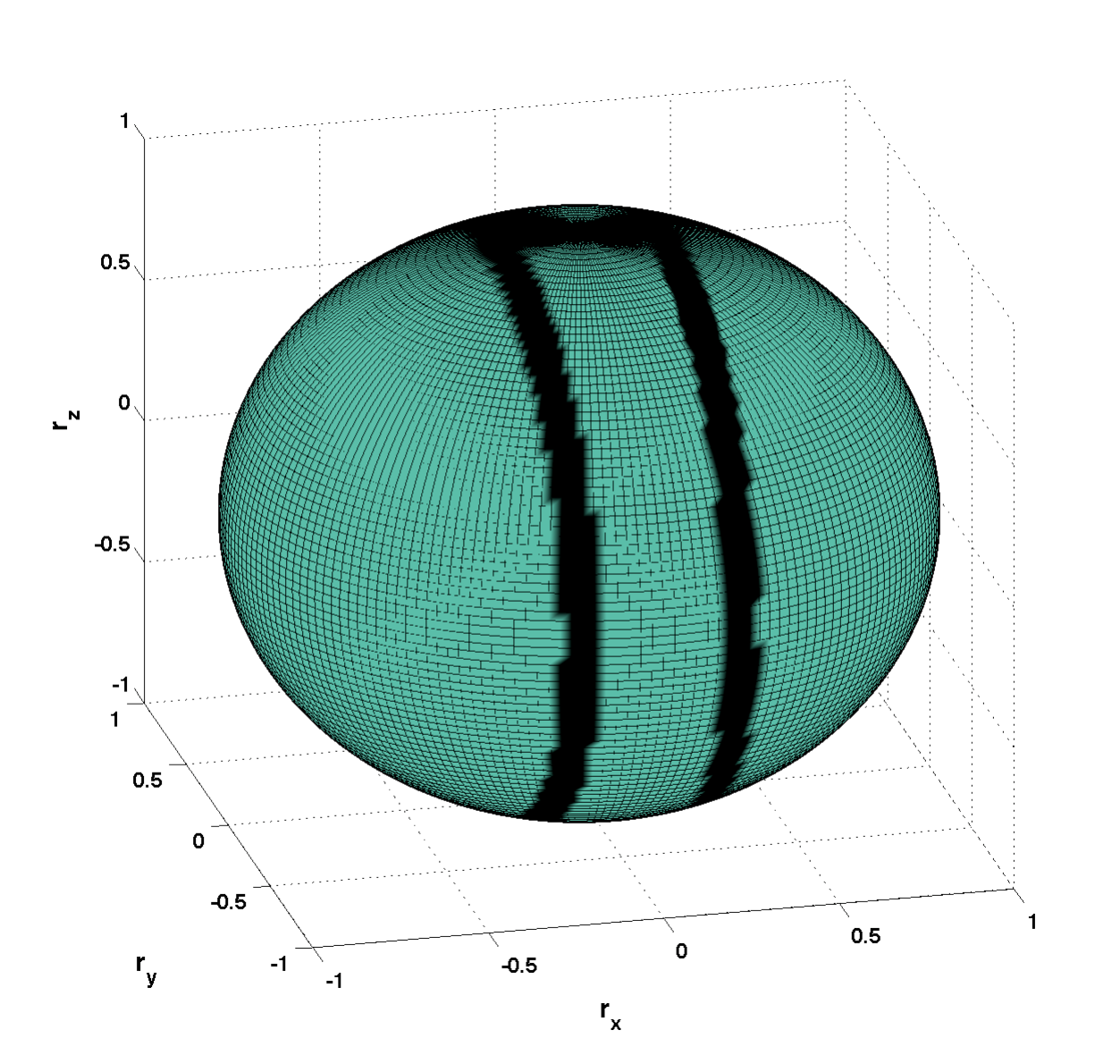}}
\subfloat[~$\mathbf{r}$-space diagram of eigenvalue types for matrix $\mathbf{A}$ when $\gamma/\Delta = \sqrt{20}$.]{ 
\label{fig:eig_phasediag:c} 
\includegraphics[width=0.25\linewidth]{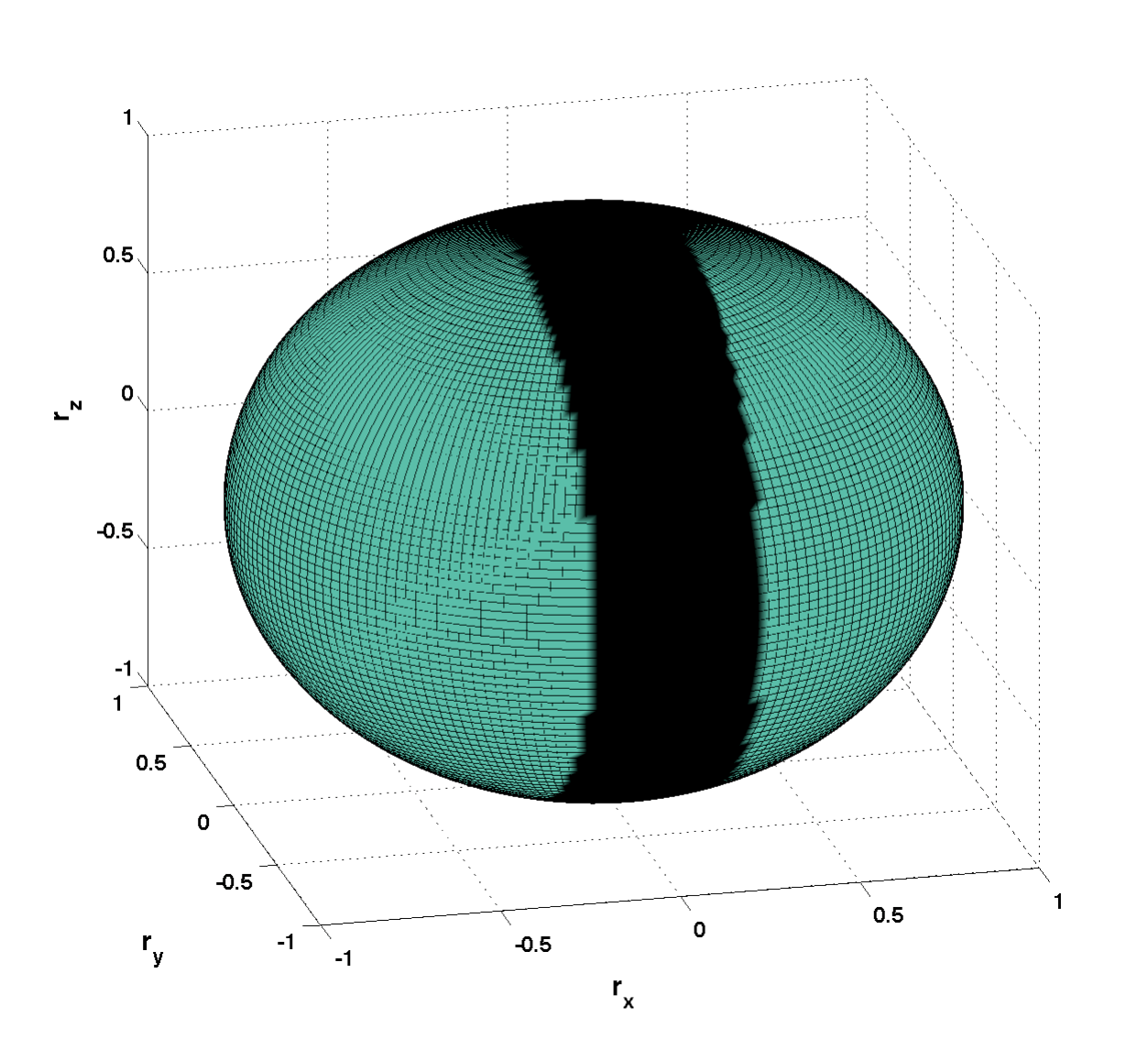}} 
\hspace{0.1\linewidth} 
\caption{Eigenvalue regions of the matrix $\mathbf{A}$ in parameter space.} 
\label{fig:eig_phasediag} 
\end{figure}

Mathematically, the change in behavior at $4\Delta $ exists because the
minus branch of the expression under the square root in equation (\ref%
{aregion}) becomes imaginary. The phase diagram in Fig. \ref%
{fig:eig_phasediag} shows the eigenvalue regimes as a function of the
parameter ratio $\gamma /\Delta $. Also, in this figure we explicitly show
the regions in $\overrightarrow{\mathbf{r}}$-space corresponding to the two
eigenvalue regimes for two specific values of $\gamma/\Delta: \sqrt{15}$ and 
$\sqrt{20}$.

From this information about the eigenvalue phase diagram, we can make the
following conclusions about the mixing behavior for various parameter
regimes:

\begin{itemize}
\item[(\textit{i})] for {$\gamma <\sqrt{12}\Delta $ we have weak
decoherence, so that for all decoherence axes (all $\overrightarrow{\mathbf{r%
}}$) we have a pair of complex conjugate eigenvalues and therefore classical mixing and the possibility of instantaneous mixing. }

\item[(\textit{ii})] for {$\gamma \geq \sqrt{12}\Delta $ and the }%
decoherence projection direction $r_{x}$ is such that $\left\vert
r_{x}\right\vert \leq \sqrt{\frac{1}{3}-\frac{\gamma ^{2}}{54\Delta ^{2}}+%
\frac{\sqrt{\left( \gamma ^{2}-12\Delta ^{2}\right) ^{3}}}{54\gamma \Delta
^{2}}}$ we have {two subcases:}

\begin{itemize}
\item[(\textit{a)}] if $\sqrt{12}\Delta \leq \gamma \leq 4\Delta $ and
decoherence projection direction $r_{x}$ also such that $\left\vert
r_{x}\right\vert \geq \sqrt{\frac{1}{3}-\frac{\gamma ^{2}}{54\Delta ^{2}}-%
\frac{\sqrt{\left( \gamma ^{2}-12\Delta ^{2}\right) ^{3}}}{54\gamma \Delta
^{2}}}$ then all eigenvalues are real, possibly repeated. There is no
oscillatory behavior in $\langle \sigma_z\rangle (t)$, and therefore no
instantaneous mixing time exists. However, the walk of course has a
classical mixing time as $\langle {\sigma _{z}}\rangle 
\xrightarrow{t\rightarrow
\infty}0$.

\item[(\textit{b})] if $\gamma >4\Delta $ all eigenvalues are real, possibly
repeated.There is no oscillatory behavior in $\langle \sigma_z\rangle (t)$,
and therefore no instantaneous mixing time exists. However, the walk of
course has a classical mixing time as $\langle {\sigma _{z}}\rangle 
\xrightarrow{t\rightarrow
\infty}0$.
\end{itemize}

\item[(\textit{iii})] for {$\gamma \geq \sqrt{12}\Delta ~$and decoherence
projection direction }$r_{x}$ is{\ such that it does not satisfy the
conditions from (\textit{ii}) and (\textit{ii-a})} then we have a pair of
complex conjugate eigenvalues and therefore classical mixing, and the possibility of instantaneous mixing.
\end{itemize}

The fact that no finite instantaneous mixing time exists when all
eigenvalues are real (for their restricted decoherence model) was
interpreted by Alagic and Russell \cite{AR05} as an analogue of the Zeno
effect where the quantum evolution of a system is hindered by its strong
interaction with an environment \cite{Mis.Sud-1977}. Following this, we
refer to the region where no instantaneous mixing time exists -- i.e. regime 
\textit{(ii}) above where all eigenvalues of $\mathbf{A}$ are real -- as the
Zeno-region, and the remainder of $(\gamma, \Delta, \overrightarrow{\mathbf{r%
}})$ parameter space as the no-Zeno region.

The presence of imaginary eigenvalues does not guarantee an instantaneous mixing time because it could still be possible that the timescale of the damping effect of the decoherence on $\expect{\sigma_{z}(t)}$ can be much faster the the characteristic period of the oscillation. We can formulate implicit conditions for the presence of instantaneous mixing by examining the expression from \erf{eq::z_t} for $\expect{\sigma_{z}(t)}$ when $\mathbf{A}$ has imaginary eigenvalues: 
\beq
\expect{\sigma_{z}(t)} = A_{1} e^{\lambda_{1}t} + A_{1}^{*} e^{\lambda_{1}^{*} t} + A_{3} e^{\lambda_{3} t}
\eeq
where we have assumed $\lambda_{1}$ and $\lambda_{2} = \lambda_{1}^{*}$ are the imaginary eigenvalues. $A_{i}$ are the time independent coefficients of the exponentials in \erf{eq::z_t}. A necesarry and sufficient condition for instantaneous mixing is: $\exists ~t ~~\textrm{s.t.}~~ \expect{\sigma_{z}(t)}\leq 0$.  Writing $A_{1}=|A_{1}|e^{i\theta}$ and $\lambda_{1}=\lambda_{R,1} + i\lambda_{I,1} - \frac{2\gamma}{3}$ and using the fact that $\lambda_{3} = -\lambda_{1}-\lambda_{1}^{*} - 2 \gamma$, this condition can be written as:
\beq
\exists t ~~ \textrm{s.t.} ~~ 2|A_{1}|e^{3\lambda_{R, 1}t}\cos(\theta + \lambda_{I, 1}t) \leq -A_{3},
\eeq
which reduces down to:
\beq
\exists t ~~ \textrm{s.t.} ~~ {|A_{1}|}e^{3\lambda_{R,1}t}\cos(\theta + \sqrt{3}\lambda_{R,1}t) \leq -2(3\lambda_{R,1}^2-4 d^2+r_z^2)
\eeq
where $A_{1}=\frac{1}{4}(4d^2+9\lambda_{R,1}^2-r_z^2)+i\frac{\sqrt{3}}{4}(4d^2-3\lambda_{R,1}^2-r_z^2(1+4\lambda_{R,1}))$ and $d=\frac{\Delta}{\gamma}$.
For any value of the physical parameters, $(\gamma, \Delta, \overrightarrow{\mathbf{r}})$, this condition can be checked for the existence of an instantaneous mixing time.

\subsection{Numerical simulations}

In this section we numerically evaluate $\langle \sigma_z \rangle (t)$ and
calculate the classical mixing time for several parameter values and $%
\varepsilon=0.001$. These simulations are summarized in Fig. \ref%
{fig:mixing_blochspheres} which show mixing time for all values of $%
\overrightarrow{\mathbf{r}}$ and three values of the physical parameter
ratio $\gamma/\Delta$. We make several observations from these plots:

\begin{itemize}
\item The mixing time can vary considerably as $\overrightarrow{\mathbf{r}}$
is varied. Although the change in mixing time is generally smooth with
changes in $\overrightarrow{\mathbf{r}}$, there are regions where the change
is abrupt (e.g. around $r_x = \pm 1$ when $\gamma/\Delta=1$). This implies
that the mixing time can potentially change drastically with the exact value
of $\overrightarrow{\mathbf{r}}$. Hence it is very important to characterize
the decoherence process accurately for determining mixing properties. A
similar conclusion was arrived at by Strauch in Ref. \cite{Strauch08} where
he demonstrates that mixing behavior differs greatly depending on the
decoherence model chosen.

\item The range of mixing times on the $\overrightarrow{\mathbf{r}}$-sphere
is smallest when $\gamma/\Delta \approx 1$. The range of mixing time
diverges when this parameter ratio is very large or very small. This
suggests an optimal $\gamma/\Delta$ parameter ratio where the interplay
between Hamiltonian dynamics and decoherence is such that decoherence in any
direction yields small mixing times. We will investigate this more
thoroughly in the next section.

\item In the Hamiltonian dominated regime, where $\gamma/\Delta<1$, we have
fast mixing near the $r_x$ axis but long mixing times in the $r_y-r_z$ plane.

\item In the decoherence dominated regime, where $\gamma/\Delta>1$, we have
fairly similar mixing times across nearly all values of $\overrightarrow{%
\mathbf{r}}$ except for a small region around $\overrightarrow{\mathbf{r}}%
=(\pm 1, 0, 0)$. The size of this region shrinks as $\gamma/\Delta$
increases, but the value of the mixing time in this region grows with the
same parameter. However, note that when the decoherence is \textit{exactly}
along the $r_x$ axis we have very short mixing times as evident from the
exact solution given by Eq. (ref{eq::Xbit-dep}) for this case.
\end{itemize}

\begin{figure}[t!]
\centering 
\subfloat[~$\gamma/\Delta = 0.01$]{ 
\label{fig:mixing_blochspheres:a} 
\includegraphics[width=0.33\linewidth]{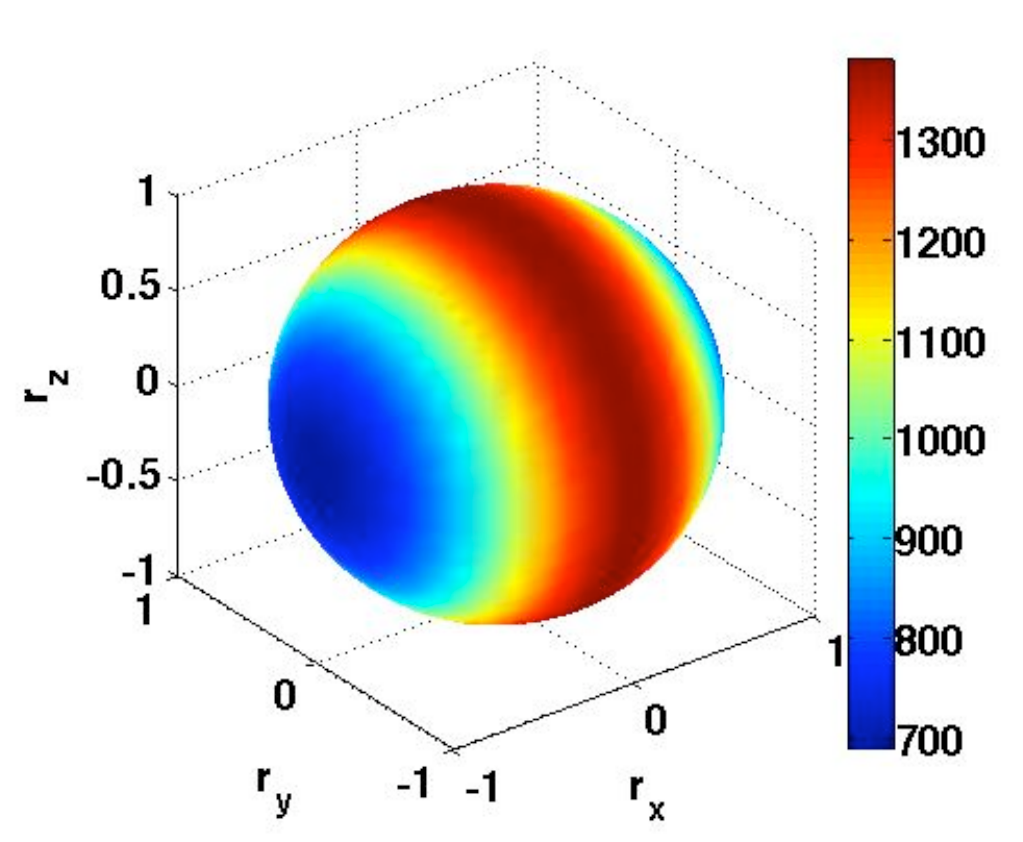}} 
\subfloat[~$\gamma/\Delta = 1$]{ 
\label{fig:mixing_blochspheres:b}
\includegraphics[width=0.33\linewidth]{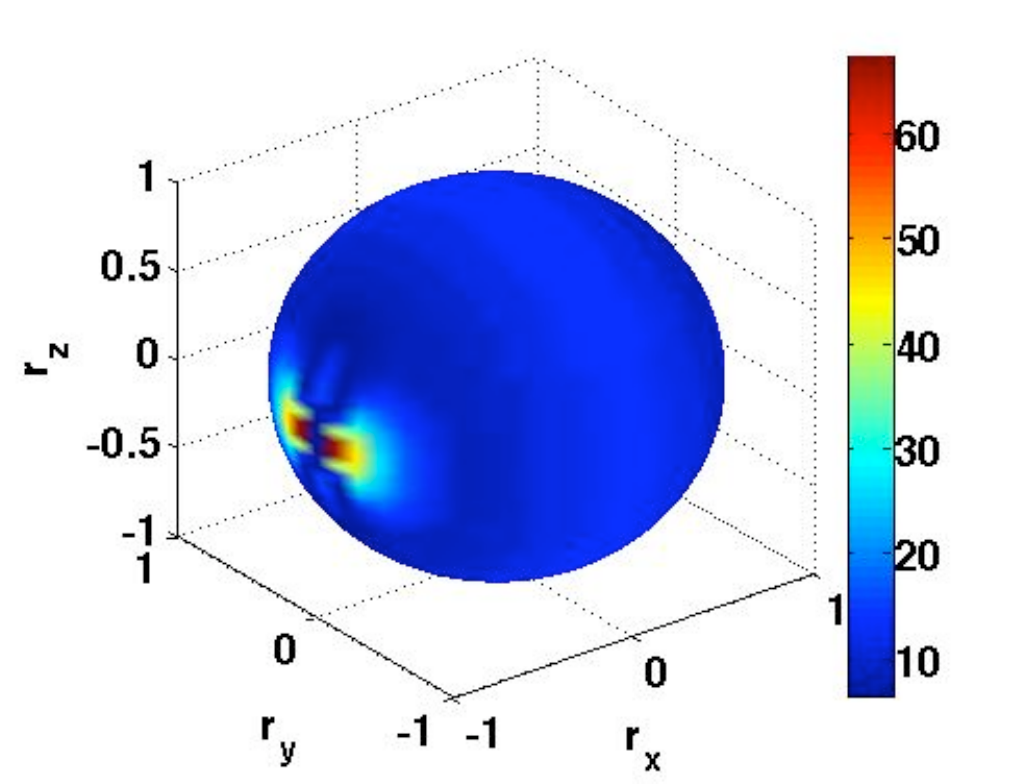}}
\subfloat[~$\gamma/\Delta = 5$]{ 
\label{fig:mixing_blochspheres:c} 
\includegraphics[width=0.33\linewidth]{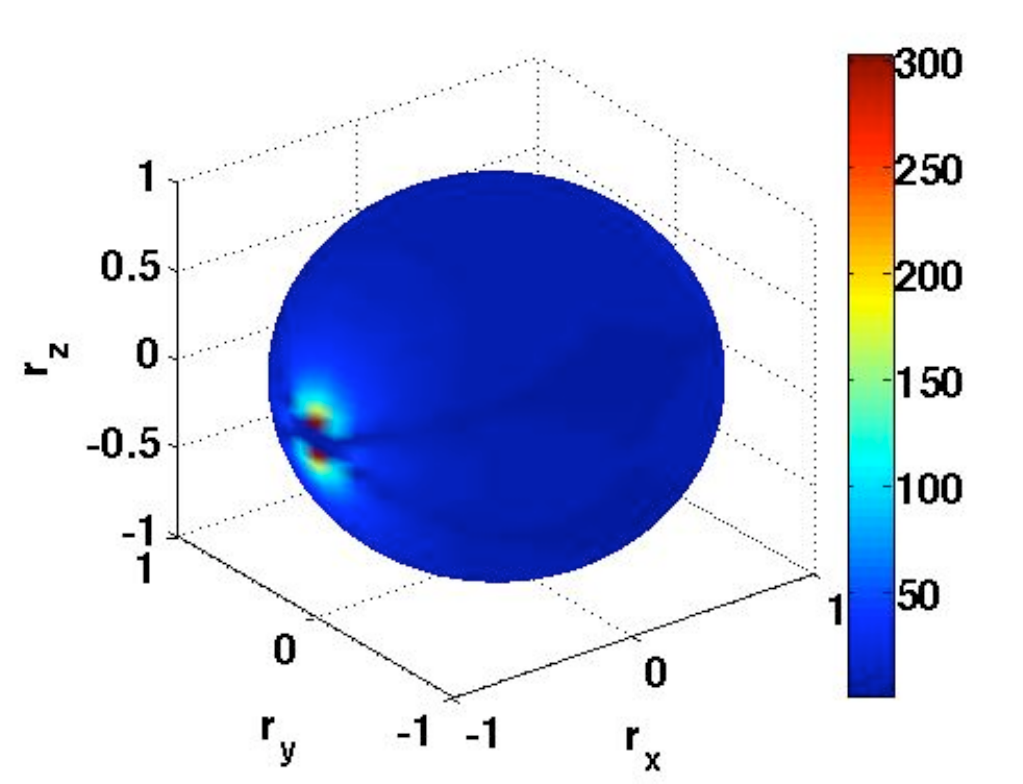}} 
\caption{Variation of single qubit mixing times with decoherence axis for several values of $\gamma/\Delta$. The color at each point on the Bloch sphere corresponds to the mixing time for single qubit dynamics when the decoherence vector is $(r_x, r_y, r_z)$. The mixing times are in units of $\Delta$ and are calculated for $\epsilon=0.001$. Note that $\langle \sigma_z \rangle (t)$ is invariant under negation of any coordinate of the decoherence axis ($r_x\rightarrow -r_x, r_y \rightarrow -r_y, r_z \rightarrow -r_z$), and so the portions of the sphere that cannot be seen can be inferred.} 
\label{fig:mixing_blochspheres} 
\end{figure}

\subsection{Optimal decoherence rate}

\begin{figure}[t!]
\centering 
\subfloat[~Characteristic behavior of mixing time versus $\gamma/\Delta$ for $\theta \in [0, \pi/2)$.]{ 
\label{fig:optmtime:a} 
\includegraphics[width=0.5\linewidth]{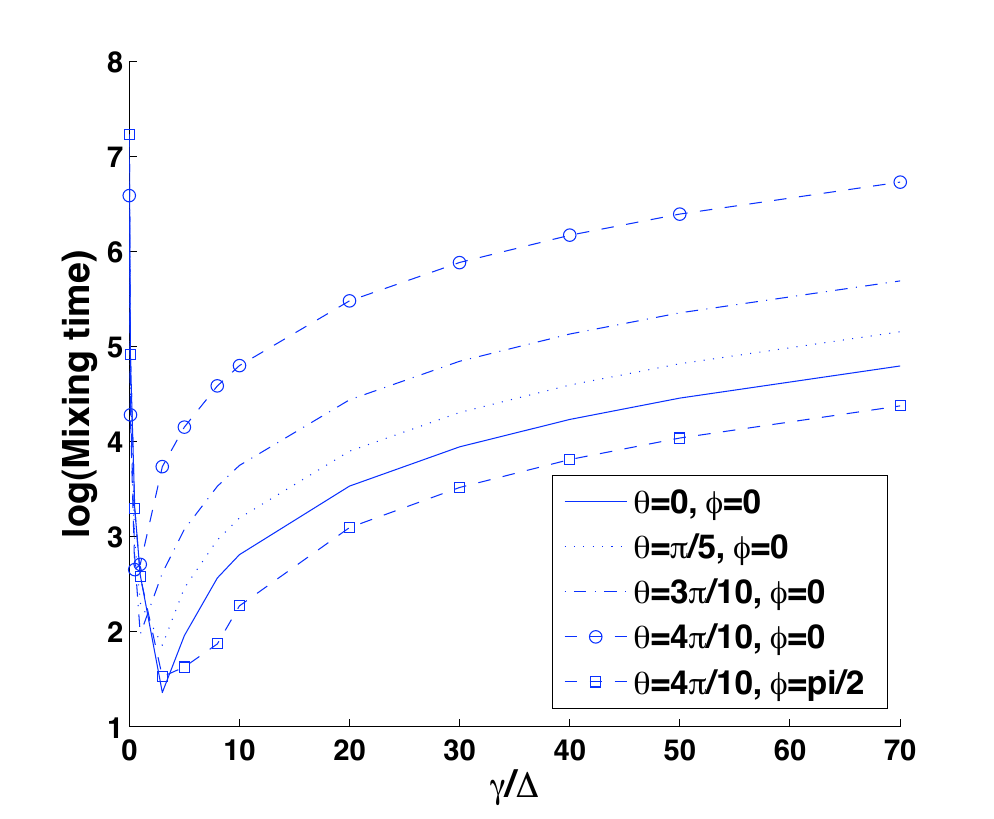}} 
\subfloat[~Characteristic behavior of mixing time versus $\gamma/\Delta$ for $\theta=\pi/2$.]{ 
\label{fig:optmtime:b}
\includegraphics[width=0.5\linewidth]{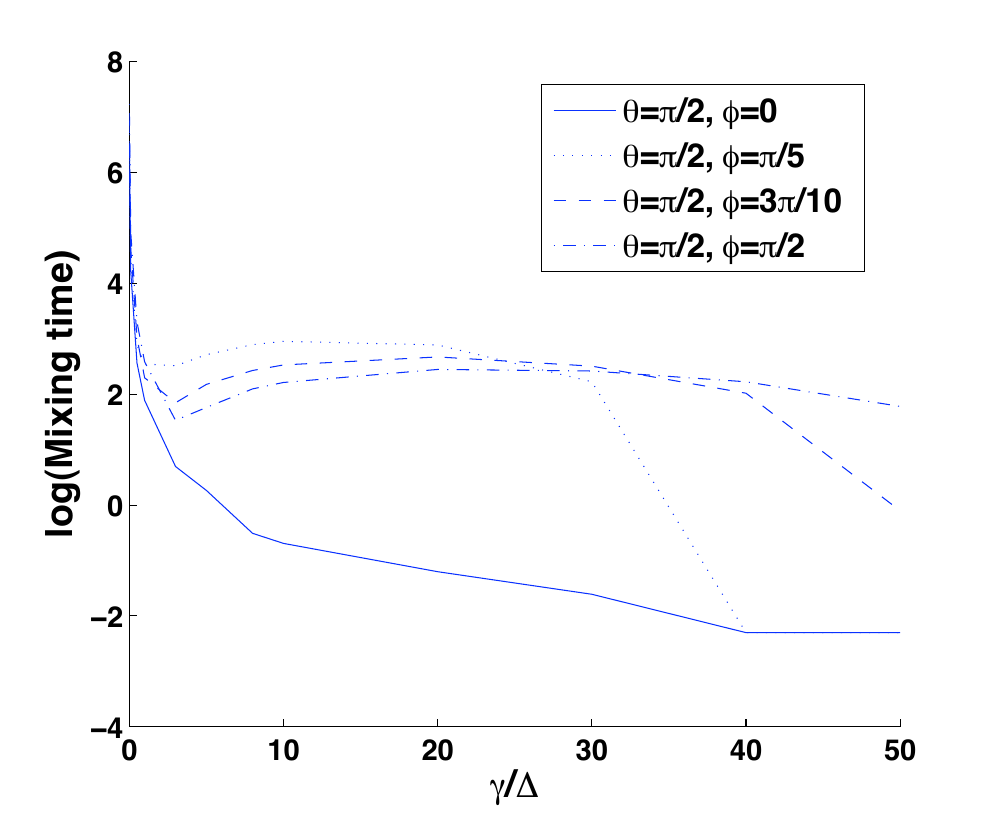}}
\caption{Scaling of (log of) single qubit mixing time with the physical parameter ratio $\gamma/\Delta$. The mixing time curves are shown for several values of $\theta$ and $\varphi$, the two angular parameters of the decoherence axis. Figure (a) shows mixing time curves for $0 \leq \theta < \pi/2$. We primarily only show curves for $\varphi=0$ in this parameter range for $\theta$ because the behavior of mixing time versus $\gamma/\Delta$ for other values of $\varphi$ is very similar: diverging mixing time as $\gamma/\Delta \rightarrow 0$ and $\gamma/\Delta \rightarrow \infty$. As $\theta \rightarrow \pi/2$ these curves show more variation with $\varphi$, but they maintain this general shape. Figure (b) shows mixing time curves for $\theta=\pi/2$ and several values of $\varphi$.} 
\label{fig:optmtime} 
\end{figure}

The numerical simulations presented in the last section suggested that the
smallest mixing time is achieved for a non-zero value of $\gamma/\Delta$.
The simulations also suggested that the optimal $\gamma/\Delta$ is around
the critical ratio $\gamma/\Delta = 1$. These observations match with Kendon
and Tregenna's conclusions in Ref. \cite{KT03} where they showed that some
amount of decoherence can lead to faster mixing of quantum walks on a line
and cycle, and faster hitting times of the quantum walk on a hypercube. Here
we fully characterize the scaling of mixing time with decoherence rate (for
the hypercube quantum walk) by numerically evaluating the mixing time for
several fixed decoherence axes. Figure \ref{fig:optmtime} shows how the
mixing time varies with $\gamma/\Delta$ for several choices of $%
\overrightarrow{\mathbf{r}}$. In general, the curves are similar for any
latitude in $\overrightarrow{\mathbf{r}}$-space, that is, for any fixed $%
\theta$. For a given $\theta$ the mixing time versus $\gamma/\Delta$ curves
(for various $\phi$) show maximum variation when $\theta=\pi/2$ (i.e. when $%
\overrightarrow{\mathbf{r}}$ is in the $x-y$ plane). Therefore we have shown
these curves on a separate plot in Fig. \ref{fig:optmtime}.

The notion of an optimal ratio $\gamma/\Delta$ is accurate for nearly all
decoherence axes. And for nearly all decoherence axes, this optimal value is
in the range $1< \gamma/\Delta < 5$. However, when $\overrightarrow{\mathbf{r%
}}$ is in the $x-y$ plane, there is no finite optimal value for $%
\gamma/\Delta$; the mixing time decreases continuously as $\gamma/\Delta$ is
increased for $\overrightarrow{\mathbf{r}}$ in the $x-y$ plane. We can gain
intuition about this result by viewing the decoherence as a localizing
phenomenon -- it tends to localize the qubit state along the axis (on its
Bloch sphere) defined by $\overrightarrow{\mathbf{r}}$ vector. And the
larger $\gamma/\Delta$ is, the faster this localization happens. For a qubit
localized in the $x-y$ plane, $\langle \sigma_z \rangle=0$ and hence a fast
localization to this plane yields fast mixing. The variation of the mixing
time when $\overrightarrow{\mathbf{r}}$ is \textit{within} the $x-y$ plane
is an interesting feature of Fig. \ref{fig:optmtime:b}. As $\overrightarrow{%
\mathbf{r}}$ approaches the $x$-axis ($\varphi=0$) the mixing time evolution
becomes closer and closer to exponential decay (with $\gamma/\Delta$).
However, away from the $x$-axis, the curves still show a local minimum
around $\gamma/\Delta \approx 1$, but the global minima are still for $%
\gamma/\Delta \rightarrow \infty$.

\section{Conclusion}

\label{sec::conc}

We complete the picture that exists in the literature for quantum random
walks on the hypercube with decoherence under the subspace projection
decoherence model. This model of decoherence is the most physically
realistic form when the quantum random walk is implemented using a register
of qubits.

The following are the important points of our work:

\begin{itemize}
\item The instantaneous mixing time property of the quantum random walk does
not necessarily disappear as the decoherence strength is increased. More
precisely, Zeno dynamics prevails only in the precisely specified regions of
the $\overrightarrow{\mathbf{r}},\gamma ,\Delta $ parameter space defined in %
\ref{subsec:eig}. Consequently, the hypercube quantum walk with decoherence
does not always limit to its classical version as the decoherence is
increased. Depending on the direction of the decoherence vector it is
possible to have oscillatory dynamics and instantaneous mixing persist as
decoherence is increased.

\item We showed numerically that, for almost all decoherence directions, a
finite decoherence rate exists for optimally fast mixing. This optimal rate
is approximately the same as the rate of Hamiltonian evolution. However, we
also showed that for certain decoherence directions ($\overrightarrow{%
\mathbf{r}}$ in the $x-y$ plane) no finite optimal decoherence rate exists
and the mixing rate increases without bound as the decoherence rate is
increased. This result is particularly relevant given recent results on
quantum walk based modeling of excitation transport in biomolecules \cite%
{Mohseni, Plenio, Rebentrost}. In these works it is argued that decoherence can lead to
faster hitting times and walker diffusion, and this is explicitly confirmed
for a simplified model of dephasing of the quantum walk. Our results suggest
that the exact model of the decoherence matters greatly, and therefore, that
an accurate model of the environmental interactions is essential to assess
the merits or demerits of decoherence to excitation transport.

\item Furthermore, we showed in section \ref{sec:examples} that randomizing the decoherence axis yields the depolarizing channel which exhibits both instantaneous and classical mixing regardless of the relative decoherence rate $\frac{\gamma }{\Delta }$. Yet another example of a quantum walk that does not decohere to a classical walk as the decoherence rate is increased. Therefore, introducing randomized decoherence may be an avenue for controlling the mixing behavior of the hypercube random walk. 
\end{itemize}

As detailed above, when implemented using a register of qubits, the hypercube random walk possesses an important symmetry that allows one to analyze its properties by simply considering the dynamics of a single qubit. Exploiting this property was crucial to deriving our results. Examining decoherence models on general quantum walks, especially ones without a lot of symmetry, is a difficult task. While we do not expect all our techniques to directly port to such situations of greater complexity, we do believe that the approach of converting the dynamics of a noisy quantum random walk into an equivalent continuous time dynamical map will be a fruitful one.

\section{Acknowledgment}

M.D. acknowledges NSF for the support under the ITR Grant No. EIA-0205641,
and thanks Umesh Vazirani and Ben Reichardt for helpful discussion.

\appendix
\section{Single Qubit Master Equation}

The master equation, (\ref{eq:master_multiqubit}), is hard to solve in
general but in our case the system Hamiltonian and decoherence operators are
a sum of the tensor products that have a special structure. Each summand is
a tensor product of elements only one of which is not the identity. 
We now show formally that this allows one to consider a combination of
single qubit evolution equations. This calculation generalizes the calculation in \cite{AR05} to deal with a decoherence axis that is along an arbitrary direction on the Bloch sphere.

\textit{Vectorization} is a technique that transforms any $n\times n$ matrix
into a $n^{2}$ dimensional vector by stacking the transposed rows of the
matrix on the top of each other. We will denote a vectorized matrix $X$ as $%
X^v$. A useful identity we will utilize involves the vectorization of the
matrix product $AXB$: $(AXB)^{v}=(B^{T}\otimes A)X^{v}$. The action of
unitary evolution on a density operator is $U_{t}\rho U_{t}^{\dagger }$, and
that is vectorized as $(U_{t}\rho U_{t}^{\dagger})^{v}=(U_{t}^{\ast }\otimes
U_{t})\rho ^{v}=\mathcal{S}_{t}\rho ^{v}$, where $\mathcal{S}_{t}$ is the
matrix form of the unitary evolution superoperator. Using this formalism it
is straight forward to derive the vectorized picture of the master equation.
Consider the discretized evolution given by Eq. (\ref{eq::weak}) with $\tau
\rightarrow dt$, and the qubit projection POVMs given by Eq. (\ref%
{eq:povm_elems}): 
\begin{equation}
\varrho _{t+dt}=(1-\gamma dt )U_{dt }\varrho _{t}U_{dt }^{\dagger }+\gamma
dt \sum_k \sum_{\alpha}M^k_\alpha[U_{dt }\varrho _{t}U_{dt}^{\dagger }]%
M_\alpha^k\,\,
\end{equation}
This can vectorized as: 
\begin{eqnarray}
\varrho _{t+dt}^{v} &=&(1-\gamma dt)\{U_{dt}\varrho _{t}U_{dt}^{\dagger
}\}^{v}+\gamma dt\mathcal{P}^{v}(U_{dt}\varrho _{t}U_{dt}^{\dagger })^{v}\,\,
\label{dtsup} \\
&=&(1-\gamma dt)(U_{dt}^{\ast }\otimes U_{dt})\varrho _{t}^{v}+ \gamma dt%
\mathcal{P}^{v}(U_{dt}^{\ast }\otimes U_{dt})\varrho _{t}^{v} \\
&=&[(1-\gamma dt)+\gamma dt\mathcal{P}^{v}][U_{dt}^{\ast }\otimes
U_{dt}]\varrho _{t}^{v}  \label{eq:vec_dynamics}
\end{eqnarray}%
where $\mathcal{P}^{v}$ is the operator:%
\begin{eqnarray*}
\mathcal{P}^{v} &=&\sum_{k=1}^n \sum_{\alpha=0}^1 [M_\alpha^{k \ast }\otimes
M_\alpha^k]\,\, \\
&=&\sum_{k=1}^n\mathbb{I}^{\otimes 2(k-1)}\otimes \lbrack \mathbb{P}%
_{0}^{\ast }(\overrightarrow{\mathbf{r}})\otimes \mathbb{P}_{0}(%
\overrightarrow{\mathbf{r}})+\mathbb{P}_{1}^{\ast }(\overrightarrow{\mathbf{r%
}})\otimes \mathbb{P}_{1}(\overrightarrow{\mathbf{r}})]\,\otimes \mathbb{I}%
^{\otimes 2(n-1-k)}
\end{eqnarray*}

This transition equation for $\varrho $ defines the system dynamics at all
times. Now, let $\varrho _{t}^{v}=\mathcal{S}_{t}\varrho _{0}^{v}$ where $%
\mathcal{S}_{t}$ is the propagator matrix for the dynamics. From Eq. (\ref%
{eq:vec_dynamics}) we know that $\varrho _{t+dt}^{v}=[(1-\gamma dt)+\gamma dt%
\mathcal{P}^{v}][U_{dt}^{\ast }\otimes U_{dt}]\mathcal{S}_{t}\varrho
_{0}^{v} $, and since this is true for any initial state $\varrho _{0}^{v}$,
we get: 
\begin{eqnarray}
\mathcal{S}_{t+dt} &=&[(1-\gamma dt)+\gamma dt\mathcal{P}^{v}][U_{dt}^{\ast
}\otimes U_{dt}]\mathcal{S}_{t}  \notag \\
&=&[(1-\gamma dt)\mathbf{1}+\gamma dt\mathcal{P}^{v}][(\mathbf{1+}%
iHdt)\otimes (\mathbf{1-}iHdt)]\mathcal{S}_{t} \\
&=&[\mathbf{1\otimes 1+}idt(H\mathbf{\otimes 1-1\otimes }H)-\gamma dt\mathbf{%
1\otimes 1+}\gamma dt\mathcal{P}^{v}]\,\mathcal{S}_{t},
\end{eqnarray}%
where we have expanded $U_{dt}=e^{-iHdt}$ to first order, and $\mathbf{1}=%
\mathbb{I}^{\otimes n}$. Taking into account that $\mathcal{\dot{S}}_{t}=%
\frac{d\,\mathcal{S}_{t}}{dt}=\lim_{dt\rightarrow 0}\frac{S_{t+dt}-S_{t}}{dt}
$ we get the differential form:

\begin{equation}
\mathcal{\dot{S}}_{t}=\left[ i[H\otimes \mathbf{1}-\mathbf{1}\otimes
H]-\gamma\mathbf{1}\otimes \mathbf{1}+\gamma\mathcal{P}^{v}\right] \,%
\mathcal{S}_{t}\,\,
\end{equation}

Using $H=\Delta \sum_{k=1}^{n}\mathbb{I}^{\otimes (k-1)}\otimes \sigma
_{x}\otimes \mathbb{I}^{\otimes (n-1-k)}$ we expand this as: 
\begin{eqnarray}
\dot{\mathcal{S}}_{t} &=&\left[ \sum_{k=1}^{n}\mathbb{I}^{\otimes
2(k-1)}\otimes \left\{ 
\begin{array}{c}
i\Delta \lbrack \sigma _{x}\otimes \mathbb{I}-\mathbb{I}\otimes \sigma
_{x}]-\gamma \mathbb{I}\otimes \mathbb{I}+ \\ 
\gamma \lbrack \mathbb{P}_{0}^{\ast }(\overrightarrow{\mathbf{r}})\otimes 
\mathbb{P}_{0}(\overrightarrow{\mathbf{r}})+\mathbb{P}_{1}^{\ast }(%
\overrightarrow{\mathbf{r}})\otimes \mathbb{P}_{1}(\overrightarrow{\mathbf{r}%
})]%
\end{array}%
\right\} \otimes \mathbb{I}^{\otimes 2(n-1-k)}\right] \mathcal{S}_{t} \\
&\equiv &\mathcal{A}\mathcal{S}_{t}.
\end{eqnarray}

Since the initial condition is $\mathcal{S}_{0}=I^{\otimes n}.$, the
solution to this differential equation is $\mathcal{S}_{t}=e^{\mathcal{A}t}$%
: 
\begin{eqnarray}
\mathcal{S}_{t} &=&\exp \left[ \sum_{k=1}^{n}\mathbb{I}^{\otimes
2(k-1)}\otimes t\left\{ 
\begin{array}{c}
i\Delta \lbrack \sigma _{x}\otimes \mathbb{I}-\mathbb{I}\otimes \sigma
_{x}]-\gamma \mathbb{I}\otimes \mathbb{I}+ \\ 
\gamma \lbrack \mathbb{P}_{0}^{\ast }(\overrightarrow{\mathbf{r}})\otimes 
\mathbb{P}_{0}(\overrightarrow{\mathbf{r}})+\mathbb{P}_{1}^{\ast }(%
\overrightarrow{\mathbf{r}})\otimes \mathbb{P}_{1}(\overrightarrow{\mathbf{r}%
})]%
\end{array}%
\right\} \otimes \mathbb{I}^{\otimes 2(n-1-k)}\right]  \\
&=&\sum_{k=1}^{n}\mathbb{I}^{\otimes 2(k-1)}\otimes \exp t\left\{ 
\begin{array}{c}
i\Delta \lbrack \sigma _{x}\otimes \mathbb{I}-\mathbb{I}\otimes \sigma
_{x}]-\gamma \mathbb{I}\otimes \mathbb{I}+ \\ 
\gamma \lbrack \mathbb{P}_{0}^{\ast }(\overrightarrow{\mathbf{r}})\otimes 
\mathbb{P}_{0}(\overrightarrow{\mathbf{r}})+\mathbb{P}_{1}^{\ast }(%
\overrightarrow{\mathbf{r}})\otimes \mathbb{P}_{1}(\overrightarrow{\mathbf{r}%
})]%
\end{array}%
\right\} \otimes \mathbb{I}^{\otimes 2(n-1-k)}  \notag \\
&=&[\exp t\left\{ i\Delta \lbrack \sigma _{x}\otimes \mathbb{I}-\mathbb{I}%
\otimes \sigma _{x}]-\gamma \mathbb{I}\otimes \mathbb{I}+\gamma \lbrack 
\mathbb{P}_{0}^{\ast }(\overrightarrow{\mathbf{r}})\otimes \mathbb{P}_{0}(%
\overrightarrow{\mathbf{r}})+\mathbb{P}_{1}^{\ast }(\overrightarrow{\mathbf{r%
}})\otimes \mathbb{P}_{1}(\overrightarrow{\mathbf{r}})]\right\} ]^{\otimes n}
\notag \\
&\equiv &[\mathcal{\bar{S}}_{t}]^{\otimes n}  \notag
\end{eqnarray}

Therefore dynamics of the system is tensor product of individual qubit
dynamics $\mathcal{\bar{S}}_{t}=e^{\bar{\mathcal{A}}t}$. The single qubit
generator can be simplified as follows:

\begin{eqnarray}
\bar{\mathcal{A}} &=&i\Delta \lbrack \sigma _{x}\otimes \mathbb{I}-\mathbb{I}%
\otimes \sigma _{x}]-\gamma \mathbb{I}\otimes \mathbb{I}+\gamma \lbrack 
\mathbb{P}_{0}^{\ast }(\overrightarrow{\mathbf{r}})\otimes \mathbb{P}_{0}(%
\overrightarrow{\mathbf{r}})+\mathbb{P}_{1}^{\ast }(\overrightarrow{\mathbf{r%
}})\otimes \mathbb{P}_{1}(\overrightarrow{\mathbf{r}})] \\
&=&i\Delta \lbrack \sigma _{x}\otimes \mathbb{I}-\mathbb{I}\otimes \sigma
_{x}]-\gamma \mathbb{I}\otimes \mathbb{I} \nn \\
&& ~~~~ + \frac{\gamma }{4}[\left( \mathbb{I}%
+\overrightarrow{\mathbf{r}}\cdot \overrightarrow{\mathbf{\sigma }}\right)
^{\ast }\otimes \left( \mathbb{I}+\overrightarrow{\mathbf{r}}\cdot 
\overrightarrow{\mathbf{\sigma }}\right) +\left( \mathbb{I}-\overrightarrow{%
\mathbf{r}}\cdot \overrightarrow{\mathbf{\sigma }}\right) ^{\ast }\otimes
\left( \mathbb{I}-\overrightarrow{\mathbf{r}}\cdot \overrightarrow{\mathbf{%
\sigma }}\right) ] \\
&=&i\Delta \lbrack \sigma _{x}\otimes \mathbb{I}-\mathbb{I}\otimes \sigma
_{x}]-\frac{\gamma }{2}\mathbb{I}\otimes \mathbb{I}+\frac{\gamma }{2}%
\overrightarrow{\mathbf{r}}\cdot \overrightarrow{\mathbf{\sigma }}^{\ast
}\otimes \overrightarrow{\mathbf{r}}\cdot \overrightarrow{\mathbf{\sigma }}.
\end{eqnarray}%
It can be easily confirmed that this is the generator for the single qubit
dynamics described by Eq. (\ref{eq::master2}) once it has been vectorized.

\section{Depolarizing Channel through the Randomized Decoherence Axis}

The generator for the single qubit dinamics when the decoherence axis is
randomized is:

\begin{equation}
\bar{\mathcal{A}}_{d}=i\Delta (\sigma _{x}\otimes \mathbb{I}-\mathbb{I}%
\otimes \sigma _{x})-\gamma \mathbb{I}\otimes \mathbb{I}+\frac{\gamma }{4\pi 
}\oint_{S^{2}}[\mathbb{P}_{0}^{\ast }(\overrightarrow{\mathbf{r}})\otimes 
\mathbb{P}_{0}(\overrightarrow{\mathbf{r}})+\mathbb{P}_{1}^{\ast }(%
\overrightarrow{\mathbf{r}})\otimes \mathbb{P}_{1}(\overrightarrow{\mathbf{r}%
})]ds\,\,\,.
\end{equation}

We can carry out this integral to get the following:

\begin{eqnarray*}
\bar{\mathcal{A}}_{d} &=&%
\begin{array}{l}
i\Delta (\sigma _{x}\otimes \mathbb{I}-\mathbb{I}\otimes \sigma _{x})-\gamma 
\mathbb{I}\otimes \mathbb{I} \\ 
+\frac{\gamma }{4\pi }\int_{\theta ,\varphi }\left[ \left( \frac{\mathbb{I}+%
\overrightarrow{\mathbf{r}}\cdot \mathbf{\sigma }}{2}\right) ^{\ast }\otimes
\left( \frac{\mathbb{I}+\overrightarrow{\mathbf{r}}\cdot \mathbf{\sigma }}{2}%
\right) +\left( \frac{\mathbb{I}-\overrightarrow{\mathbf{r}}\cdot \mathbf{%
\sigma }}{2}\right) ^{\ast }\otimes \left( \frac{\mathbb{I}-\overrightarrow{%
\mathbf{r}}\cdot \mathbf{\sigma }}{2}\right) \right] d\overrightarrow{%
\mathbf{r}}%
\end{array}
\\
&=&%
\begin{array}{l}
i\Delta (\sigma _{x}\otimes \mathbb{I}-\mathbb{I}\otimes \sigma _{x})-\gamma 
\mathbb{I}\otimes \mathbb{I} \\ 
+\frac{\gamma }{8\pi }\int_{\theta ,\varphi }(\mathbb{I}\otimes \mathbb{I}+%
\overrightarrow{\mathbf{r}}\cdot \overrightarrow{\mathbf{\sigma }}^{\ast
}\otimes \overrightarrow{\mathbf{r}}\cdot \overrightarrow{\mathbf{\sigma }})d%
\overrightarrow{\mathbf{r}}%
\end{array}
\\
&=&i\Delta (\sigma _{x}\otimes \mathbb{I}-\mathbb{I}\otimes \sigma _{x})-%
\frac{\gamma }{2}\mathbb{I}\otimes \mathbb{I}+\frac{\gamma }{6}(\sigma
_{x}\otimes \sigma _{x}+\sigma _{y}^{\ast }\otimes \sigma _{y}+\sigma
_{z}\otimes \sigma _{z}).
\end{eqnarray*}%
where we have used $r_{x}=\sin \theta \cos \varphi ,$ $r_{y}=\sin \theta
\sin \varphi ,$ $r_{z}=\cos \theta .$ This leads to the solution for the
single qubit dynamics:

\begin{equation}
\rho ^{v}\left( t\right) =e^{\bar{\mathcal{A}}_{d}t}\rho ^{v}\left( 0\right)
\,.  \label{sev}
\end{equation}

Changing the basis for $\bar{\mathcal{A}}_{d}$ to the eigenbasis basis,
exponentiating, and returning to the original basis we get: 
\begin{equation}
\rho \left( t\right) =\frac{1}{2}\left( 
\begin{array}{cc}
1+e^{-\frac{2\gamma t}{3}}\cos (2\Delta t) & -ie^{-\frac{2\gamma t}{3}}\sin
(2\Delta t) \\ 
ie^{-\frac{2\gamma t}{3}}\sin (2\Delta t) & 1-e^{-\frac{2\gamma t}{3}}\cos
(2\Delta t)%
\end{array}%
\right) \,\,\,\,.
\end{equation}
with $\rho(0) = |0\rangle\langle 0 |$. The value of $\rho \left( t\right) _{00}$ and $\rho \left( t\right) _{11}$
determines the probability of measurement in basis $\{{|{0}\rangle ,|{1}%
\rangle \}.}$ The eigenvalues of our operator $\sigma (\bar{\mathcal{A}}%
_{d})=\{0,-\frac{2\Delta \gamma t}{3},-\frac{2\Delta t}{3}(\gamma -3i),$ $%
\frac{2\Delta t}{3}(\gamma +3i)\}$ determine the probability distribution.
The expressions for $\rho \left( t\right) _{00}$ and $\rho \left( t\right)
_{11}$ show that \textit{regardless} of the rate of decoherence
instantaneous mixing exists.


\end{document}